\documentclass[epj]{svjour}
\usepackage[utf8]{inputenc}
\def\bea {\begin{eqnarray}}
\def\eea {\end{eqnarray}}
\usepackage{multirow}
\usepackage{adjustbox}
\usepackage{rotating}
\usepackage{varwidth}
\usepackage{color}
\usepackage{array}
\usepackage{amsmath}
\usepackage{amssymb}
\usepackage{graphics}
\usepackage{graphicx}
\usepackage{mathrsfs}

\newcommand{\bl}{\color{black}}  
  
\begin{document}
\onecolumn
\title{Constraining nuclear matter parameters from correlation systematics:
a  mean-field perspective }
\author{B. K. Agrawal\inst{1,2}\fnmsep\thanks{\email{sinp.bijay@gmail.com}} \and
Tuhin Malik \inst{3} \and J.N. De \inst{1} \and S. K. Samaddar \inst{1} }
\institute{Saha Institute of Nuclear Physics, 1/AF
 Bidhannagar, Kolkata 700064, India. \and Homi Bhabha National Institute,
 Anushakti Nagar, Mumbai 400094, India. \and Department of Physics,
 BITS-Pilani, K. K. Birla Goa Campus, Goa 403726, India}

\abstract{
The nuclear matter parameters define the nuclear equation
of state (EoS), they appear as coefficients of expansion around the
saturation density of symmetric and asymmetric nuclear matter. We review
their correlations with several properties of finite nuclei and of neutron
stars within mean-field frameworks.  The lower order nuclear matter
parameters such as the binding energy per nucleon, incompressibility
and the symmetry energy coefficients are found  to be constrained in
narrow limits through their strong ties with selective properties of
finite nuclei.  From the correlations of  nuclear matter parameters with
neutron star observables, we further review how precision knowledge of
the radii and tidal deformability of neutron stars in the mass range $1
- 2 M_\odot$ may help cast them in narrower bounds.  The higher order
parameters such as the density slope and the curvature of the symmetry
energy or the skewness of the symmetric nuclear matter EoS are, however,
plagued with larger uncertainty. From inter-correlation of these higher
order nuclear  matter  parameters with lower order ones, we explore how
they can be brought to more harmonious bounds.  }

%
\maketitle
\section{Introduction}
\label{intro}
Precise determination of the equation of state (EoS) of nuclear matter
is one of the major goals  in nuclear physics. Relying on a realistic
nucleonic interaction, in a microscopic framework, the energy density
$\mathcal{E}$ of the system (a finite nucleus or a macroscopic nuclear
system) is computed; the parameters of the interaction or of the energy
density functional (EDF) are tuned so that the predicted observables
calculated with the EDF match with the experimental data. In a broad
sweep, the EoS or the EDF then entails knowledge of the diverse nuclear matter 
parameters  that define infinite nuclear matter: its saturation density
$\rho_0$, the energy per nucleon $e(\rho_0)$, the incompressibility
$K(\rho_0)$, the symmetry energy coefficient $C_2(\rho_0)$ and their
density derivatives of different orders. These nuclear matter  parameters  enter
into the EoS as Taylor expansion coefficients around the saturation
density; when precisely determined they stand out as irreducible elements
of physical reality as they have the mark of defining a model-independent
nuclear  EoS.

Not all of the  nuclear matter  parameters  are known in very good bounds. The profusion
of data on the masses of atomic nuclei has kept the uncertainties in the
values of $\rho_0$ and $e(\rho_0)$ quite small 
\cite{Myers1969,Myers1980,Moeller2012}. Correlation systematics has proved
to be a useful tool in arriving at values of many others. For example, the
centroid energy $E_{\scriptsize{GMR}}$ of the isoscalar giant monopole resonance (ISGMR) 
is a measure of $K(\rho_0) (= K_0 = 9\rho_0^2\frac{\partial^2 e}{\partial
\rho^2}|_{\rho_0}$). The correlation diagram drawn between the predicted ISGMR
energies for a heavy nucleus like $^{208}$Pb with different EDFs against
the related nuclear matter  parameter  $K_0$ pertaining to the EDFs has proved
to be an effective method of projecting its value \cite{Blaizot1980}
at saturation density. Noticing the recently found remarkable soft
nuclear matter  compression from the ISGMR data in Sn and Cd isotopes
\cite{Lui2004,Li2007,Garg2007,Li2010,Patel2012},
questions are asked \cite{Khan2010,Khan2012,Khan2013} on whether
ISGMR energy is a reflection of compression at saturation density or
whether $E_{\scriptsize{GMR}}$ is related to the nuclear matter  incompressibility averaged over 
the whole density range within the nucleus. The correlation
systematics was, however, not severely called into question. In several
non-relativistic and relativistic EDFs, the ISGMR energy $E_{\scriptsize{GMR}}$ was
found to be well correlated  to $M(\rho_c))$, the density derivative
of $K(\rho)~[M(\rho_c)  = 3\rho_c K'(\rho_c)]$ at a crossing  density $\rho_c$
that is close to the average density in a nucleus \cite{Khan2012}. A
subtle correlation analysis at the end then leads to a value of $K_0$
\cite{De2015}, $M_c$ and also to $K_c$, the incompressibility at $\rho_c$
\cite{Khan2012}.

There is a cultivated focus in recent times in tightening the bounds
on the values of the isovector nuclear matter  parameters, namely, the
symmetry energy $C_2^0(\equiv C_2(\rho_0))$ , its density derivative $L_0
(\equiv L(\rho_0))$, the symmetry incompressibility $K_{\rm sym}^0 (\equiv
K_{\rm sym}(\rho_0))$ etc \cite{RocaMaza2018}.  They have a fundamental
role in deciding pressures in neutron-rich matter; they determine the
nuclear masses, the neutron-skin thickness and the size and structure of
neutron stars. From a Bethe-Weizacker type of expansion of the nuclear
binding energy in powers of the  mass number  $A$, the symmetry energy
$C_2^0$ is well obtained \cite{Myers1969,Moeller2012}; the exhibited
compact correlation of the double differences of the 'experimental
symmetry energies' of four neighboring nuclei with the nuclear mass
number $A$ yields \cite{Jiang2012}, however, the value of the volume and
surface symmetry energies at $\rho_0$ with much less uncertainty. No less
important is the strong correlation of the value of the centroid of the
isovector giant dipole resonance (IVGDR) energy in spherical nuclei with
the symmetry energy $C_2(\rho)$ \cite{Trippa2008} at $\rho \sim 0.1$fm
$^{-3}$ found in Skyrme  EDFs in imposing a quantitative  constraint on
the symmetry energy at a sub-saturation density. The value ($C_2(\rho
\sim 0.1)\approx 23.3\pm 0.8$ MeV) is in extremely good agreement with
that found from analysis of experimental isoscalar and isovector giant
quadrupole resonances of the nucleus $^{208}$Pb \cite{RocaMaza2013}.

Systematics in relativistic and non-relativistic models have
revealed that there is a strong correlation between the density
derivative of symmetry energy $L_0$ with the neutron-skin thickness
$\Delta r_{np}$ [$=(R_n -R_p)$, $R_n$ and $R_p$ are the neutron and
proton root mean squared radius (rms) ] of a heavy nucleus like
$^{208}$Pb \cite{Brown2000,Typel2001,Centelles2009}. Analyzing
the correlation systematics of nuclear isospin with neutron-skin
thickness for a series of nuclei in the ambit of the droplet model,
the Barcelona group \cite{Centelles2009,Warda2009} predicted $L_0$ in a
comparatively narrow window, but it suffers from the unavoidable strong
interaction-related uncertainties in the neutron-skin thickness derived
from anti-protonic atom experiments. Finding a model-independent precise
value of the neutron-skin thickness is a major  challenge  still not
quite accomplished \cite{Abrahamyan2012}, so the converse route of
finding $L_0$ and then predicting $\Delta r_{np}$ from correlation
systematics has been taken by many. The often shifting values of
$L_0$ obtained from different observables like pygmy dipole resonance
\cite{Carbone2010}, isovector giant dipole resonance \cite{Trippa2008},
nucleonic emission ratios \cite{Famiano2006}, nuclear masses
\cite{Moeller2012,Agrawal2012,Agrawal2013} or even astrophysical inputs of
neutron star radii \cite{Steiner2012} render $L_0$ and thus
$\Delta r_{np}$ still somewhat uncertain. Of late, co-variance analysis
with masses of selective highly asymmetric nuclei \cite{Mondal2015}
and experimental data on collective isovector excitations in nuclei
\cite{Paar2014} tend to constrain $L_0$ tightly, the constraint, however,
depends much on the precision of the relevant experimental data.

The multitude of EDFs are rooted to various sets of selective experimental
data that may have occasional overlaps. It is therefore not surprising
that the bulk nuclear matter  parameters  (tied to the various parameters
in the EDFs) may be intercorrelated and so the nuclear observables
display built-in correlations with the nuclear matter  parameters  pertaining to
the concerned EDFs.  As examples, further to those mentioned earlier,
the core-crust transition density $\rho_t$ in neutron star is found to
be correlated \cite{Horowitz2001} to the neutron-skin $\Delta r_{np}$
in the nucleus $^{208}$Pb, $\rho_t$ is also seen to be correlated with
the symmetry density derivative $L_0$ \cite{Horowitz2001,Ducoin2011},
$\Delta r_{np}$ has a correlation with the product of nuclear dipole
polarizability $\alpha_D$ with the symmetry energy $C_2^0$
\cite{RocaMaza2015}.  Knowledge of a better known entity then throws
light on the one lesser known, this is the kernel of the correlation
systematics.

{\bl The observable properties of neutron stars offer fresh grounds for
exploring the nuclear matter  EoS on a large density plane, spanning
a few times the nuclear saturation density.  Behind the solid crust
of $\sim 1$ km thickness lies the central homogeneous liquid core,
its density increases as one approaches the center.}  The outer crust is
inhomogeneous with neutron-rich nucleons and degenerate electrons. The
inner crust, loosely speaking, starts when with increasing density
and pressure, neutronization sets in leading to the existence of a
neutron ocean with inhomogeneous nucleonic clusters and electrons.
The structure of the inner crust is modeled as a lattice in a body
centered cubic formation  with the electron gas circulating throughout
the structure \cite{Teukolsky1983}, the free neutrons are assumed to
have no effect. However, if the interactions between the neutrons and
the lattice are accounted for, a rethink on the structure of the crust
and its response to perturbations may be needed \cite{Kobyakov2014}. The
density at which neutrons drip off from nuclei is rather well known,
but the transition density at the inner edge separating the solid crust
from the liquid core is still not well settled.  The bottom layer of the
inner crust consists mostly of exotic nuclear structures collectively
known as nuclear pasta \cite{Lorenz1993,Berry2016}. The transition
density offers important insights into the origin of pulsar glitches
from its relation to the crustal fraction of the moment of inertia of
a neutron star \cite{Madhuri2017} with additional information on the
gravitational wave radiation \cite{GuerraChaves2019} by the neutron star
when subjected to a very intense gravitational field.

In this article, we do not deal with the EoS of the neutron star crust,
it is assumed to be known \cite{Baym1971}, we rather focus on the nuclear matter 
EoS for the homogeneous core region, from around the saturation density
$\rho_0$ to the central density of around $\sim 5 -6\rho_0$, and on the
specifics of the nuclear matter  parameters  that describe them. The analysis of
directed and elliptic flow \cite{Danielewicz2002} and kaon production
\cite{Fuchs2006,Fantina2014} from heavy ion collisions (HIC) has helped
understand the nuclear matter  EoS at supra-normal densities. Recently, the
detection of gravitational waves from the GW170817 binary neutron star
merger \cite{Abbott2017} has added much impetus to effectively map the
EoS at densities relevant to neutron stars.

Microscopic analysis of different kinds of laboratory data, on the other
hand, have probed the nuclear matter  EoS at around the saturation density with
different levels of satisfaction. Attempts have been made to
find a link of these two \cite{Malik2018,Malik2019,Tsang2019} through the
nuclear matter  parameters , the common denominators appearing as Taylor expansion
coefficients in the nuclear matter  EoS.  Scanning the multitude of terrestrial
and astronomical data, coherent systematics have been followed to arrive
at their values as best as possible, we aim to describe
part of it in this article.

An alternate route, based on ab-initio models (for a brief review,
see \cite{Ekstroem2019}) aims at analyzing the properties of finite
nuclei and of infinite nuclear matter on a more fundamental
level. With a multitude of theoretical descriptions of the
interaction between nucleons at various levels of phenomenology
\cite{Machleidt2001,Entem2003,Kolck1999,Epelbaum2009,Machleidt2011},
significant efforts have been employed in solving the many-body
Schr\"odinger equation \cite{Lee2009,Barrett2013,Carbone2013,Hagen2014}
with as few uncontrolled approximations as possible. The curse of
dimensionality in solving the complex many-body  problem along with the
somewhat hazy knowledge of the nucleonic interaction has limited their
predictive power upto only light nuclei and upto density close to
saturation density and somewhat beyond
\cite{Hergert2016,Morris2018,Holt2019,Sammarruca2015}, that too
with a precision that may be insufficient. A complementary approach was
recently proposed aimed explicitly in bridging the ab-initio methods
with an ab-initio equivalent Skyrme EDF \cite{Salvioni2020}. This was
found to describe properties of nuclei and nuclear matter poorly. Though
fundamental, we do not explore the quantification of the nuclear matter  parameters 
in the framework of ab-initio models, but settle on the nuclear matter  EoS
based on the present broad knowledge  of the effective nucleon-nucleon
interaction aided by a mean-field perspective.

\section{Theoretical Inputs}
The energy per nucleon of asymmetric homogeneous nuclear matter at density $\rho$
can be written as, 
\bea
\label{xeq1}
  e(\rho,\delta)=e(\rho,\delta=0)+C_2(\rho) \delta^2+C_4(\rho)\delta^4+ \cdots,
\eea
where $\delta=(\rho_n-\rho_p)/\rho$ is the isospin asymmetry of the
system.  The first term on the right hand side (r.h.s) of Eq. (\ref{xeq1})
is the energy corresponding to  symmetric nuclear matter (SNM),
the other terms are the asymmetry contributions. The coefficients
$C_n(\rho)$ are collectively called the symmetry coefficients. 
For nearly all energy density functionals,
it is found that the parabolic approximation (terms upto $C_2$ in
Eq. (\ref{xeq1})) is quite reasonable for densities upto $\sim 1.5 \rho_0$
\cite{Chen2009,Constantinou2014,Vidana2002,GonzalezBoquera2017}.  Around the
saturation density $\rho_0$, the SNM component $e(\rho,\delta=0)$ can
be expanded as,

\bea
\label{xeq3}
e(\rho,\delta=0)=e_0+\frac{1}{2}K_0 \epsilon^2 + \frac{1}{6} Q_0 \epsilon^3 
+\frac{1}{24}Z_0\epsilon^4+ \cdots,
\eea

where $e_0=e(\rho_0,\delta=0)$, $\epsilon=(\rho-\rho_0)/3 \rho_0$, $K_0$
$(=9 \rho_0^2 \frac{\partial^2 e}{\partial \rho^2}|_{\rho_0})$ is the
isoscalar incompressibility, $Q_0(=27 \rho_0^3 \frac{\partial^3 e}{\partial
\rho^3}|_{\rho_0})$ the skewness parameter etc. The parameter $Q_0$ is
related to the density derivative of the nuclear matter  incompressibility $(M=3
\rho \frac{d K}{d \rho})$ at $\rho_0$, $M_0=12 K_0 + Q_0$.  The symmetry
coefficient $C_2(\rho)$ can likewise be expanded as,

\bea
\label{xeq4}
C_2(\rho)=C_2(\rho_0)+ L_0 \epsilon + \frac{1}{2} K_{\rm sym}^0 \epsilon^2 + \frac{1}{6}
Q_{\rm sym}^0 \epsilon^3 +\frac{1}{24}Z_{\rm sym}^0\epsilon^4+ \cdots,
\eea

where $C_2(\rho_0) $ is traditionally taken to be the symmetry energy coefficient
of nuclear matter, $L_0(=3 \rho_0 \frac{\partial C_2}{\partial
\rho}|_{\rho_0})$ is the symmetry slope, $K_{\rm sym}^0(= 9 \rho_0^2
\frac{\partial ^2 C_2}{\partial \rho^2}|_{\rho_0})$ the symmetry
curvature or symmetry incompressibility, $Q_{\rm sym}^0 (=27 \rho_0^3
\frac{\partial ^3 C_2 }{\partial \rho^3}|_{\rho_0})$ the symmetry
skewness coefficient etc. Precise values of these nuclear matter  parameters
entering in Eq.(\ref{xeq3})and (\ref{xeq4}) determine the nuclear matter  EoS
in a model independent way. Sophisticated analysis of laboratory data
with microscopic theoretical tools helps to find values for some of
these quantities. The lower order density derivatives are particularly
known in reasonable bounds.  The higher density derivatives like $Q_0,
K_{\rm sym}^0$ etc. are somewhat uncertain. The lower and higher density
derivatives may, however, be linked in a correlated chain, a correlation
analysis then helps to keep the uncertain nuclear matter  parameters in somewhat
stringent  bounds.

\section{Symmetric nuclear matter}
In a thermodynamic system, the Euler equation relates  the energy per
particle with its chemical potential and the pressure of the system.
In this section, we try to build a simple EDF exploiting this relation.
We focus first on the one component system, the SNM.  The correlation
between the lower and the  higher order density derivatives of energy
is  manifested here in a simple manner  without loss of generality.

\subsection{ EDF: a thermodynamic view point}
The Euler equation reads as

          \bea
          \label{xeq5}
            \mu=e+\frac{P}{\rho},
          \eea
          where $\mu$ is the chemical potential of a nucleon in SNM, $e$ its energy,
          $P$ the pressure of the system, all at density $\rho$, at a temperature
          $T=0$. The chemical potential equals the single particle energy $\varepsilon_F$ 
          at the Fermi surface, 
          \bea
          \label{xeq6}
          \mu \equiv \varepsilon_F = \frac{P_F^2}{2 m}+ U= \frac{P_F^2}{2 m^{\star}} + V.
          \eea
          Here $P_F$ is the Fermi momentum and 
          $m^{\star}$ the  effective nucleon mass given by
$\hbar^2/2m^{\star}=\delta {\mathcal{H}}/\delta
          \mathcal{K}$.  The single particle potential $V$ is calculated
          from $V=\delta \mathcal{H}/\delta \rho$.  Here $\delta $ refers
          to the functional derivative, $\mathcal{H}$ is the energy
          density, $\frac{\hbar^2}{2 m} \mathcal{K}$ is the kinetic
          energy density
           and $m$ the bare nucleon mass. The single particle
          potential can be redefined as $U$ by including within it the
          effective mass contribution as seen  in Eq. (\ref{xeq6}). No
          special assumption is made about the nucleonic interaction
          except that it is density dependent to simulate many body
          forces and that it depends quadratically on the momentum. Then,
          the single-particle potential $U$ separates into three parts,

          \bea 
 	\label{xeq7}
              U= V_0 + P_F^2 V_1 + V_2.
          \eea
The term $(V_0+P_F^2 V_1)$ is the Hartree-Fock potential and the last term
is the rearrangement potential that arises from the density dependence
in the interaction. The term $V_1$ is the result of momentum dependence
of the interaction:

           \bea \label{xeq8} \frac{P_F^2}{2 m^{\star}}= \frac{P_F^2}{2 m}
           + P_F^2 V_1,
              \eea
           so that, \bea \label{xeq9} \frac{1}{m^{\star}}=\frac{1}{m}+
           2 V_1.  \eea In general, the effective mass is momentum
           and energy dependent; in the mean-field level, the energy
           dependence is ignored and the momentum dependence is
           taken at the Fermi surface. The rearrangement term
           does not enter explicitly in the energy expression
           when written in terms of the mean-field potential
           \cite{Brueckner1959,Bandyopadhyay1990}.

        The energy per nucleon at density $\rho$ is given by \bea
        \label{xeq10} e &=& <\frac{p^2}{2m}> + \frac{1}{2}(<p^2>V_1+V_0)
        \nonumber \\
          &=& \frac{1}{2}(1 + \frac{m^{\star}}{m})<\frac{p^2}{2
          m^{\star}}> + \frac{1}{2}V_0.
        \eea Using Eqs (\ref{xeq5},\ref{xeq6},\ref{xeq7},\ref{xeq10}),
        this is written as, \bea \label{xeq11}
         e=\frac{P_F^2}{10 m}[3 - 2 \frac{m}{m^{\star}}] - V_2
         +\frac{P}{\rho},
        \eea where $<p^2>=\frac{3}{5}P_F^2$ is taken.The effective
        mass is density dependent, to lowest order, it is taken as
        $\frac{m}{m^{\star}}=1+\frac{k_+}{2}\rho$ \cite{Bohr1975}, the
        rearrangement potential $V_2= a \rho^{\tilde \alpha}$ emerges for
        finite range density dependent forces \cite{Bandyopadhyay1990}
        or for Skyrme  interactions, and so we retain this form. The
        quantities $a,k_+$ and $\tilde \alpha$ are numbers.

        Writing for $\frac{P_F^2}{2 m }= \gamma \rho^{2/3}$
        ($\gamma=(\frac{3}{2} \pi^2)^{2/3} \hbar^2/2m $), the energy of
        a nucleon in SNM is then \bea \label{xeq12}
           e(\rho)=\frac{\gamma}{5} \rho^{2 / 3}[1-k_+ \rho]-a
           \rho^{\tilde \alpha}+\frac{P}{\rho}.
        \eea Since $P=\rho^2 \partial e/\partial \rho$, from
        Eq. (\ref{xeq12}),

        \bea
        \label{xeq13}
        P(\rho)=\frac{\gamma}{15} \rho^{5 / 3}-\frac{\gamma}{6} 
        k_+\rho^{8 / 3}-\frac{1}{2} \tilde \alpha a \rho^{\tilde \alpha+1}+\frac{\rho}{18} K(\rho),
        \eea
        where use has been made of the relation for incompressibility  $K= 9 \frac{dP}{d\rho}$. 
Successive density derivatives of
        Eq. (\ref{xeq13}) give iteratively a relation of a 
   lower order  density derivative 
        with a higher order density derivative, like:
        \bea
        \label{xeq14}
        K(\rho)&=&2 \gamma \rho^{2 / 3}-8 \gamma k_{+} \rho^{5 / 3}-9 {\tilde \alpha}({\tilde \alpha}+1) a {\rho}^{\tilde
\alpha}+\frac{M(\rho)}{3},
\eea where,  
\bea
      M(\rho)&=&3 \rho \frac{d K}{d \rho}=27 \rho \frac{\partial^{2} P}{\partial \rho^{2}}. \label{xeqmc} 
        \eea
        
        At the saturation density $\rho_0$, $P=0$. From 
Eq. (\ref{xeq12}) and Eq(\ref{xeq13}), one then gets 
        \bea
        \label{xeq15}
        e_{0}=\frac{\gamma}{5} \rho_{0}^{2 / 3}\left[1-k_{+} \rho_0 \right]-a \rho_{0}^{\tilde \alpha},
        \eea
        and
        \bea
        \label{xeq16}
        \frac{1}{2} \tilde\alpha a \rho_0^{\tilde \alpha}+\frac{\gamma}{6} k_{+} 
        \rho_{0}^{5 / 3}-\left(\frac{K_{0}}{18}+\frac{\gamma}{15} \rho_{0}^{2 / 3}\right)=0.
        \eea
        
        If the value of the  effective mass $m_0^\star$
($\equiv m^{\star}(\rho_0)$) is given,
then $k_+$ is known. If $e_0$ and  the incompressibility 
        $K_0 (\equiv K(\rho_0))$ are further assumed to be known,
        then the energy of SNM around $\rho_0$ can be calculated from
        Eq. (\ref{xeq3}) as the values of $Q_0$ and
further terms involving the higher density
        derivatives are related in the correlation
chain (Eq. (\ref{xeq15}) and (\ref{xeq16}) can be solved for
        the unknown quantities $a$ and $\tilde \alpha$ ) \cite{De2015}.
From given values of $e_0, \rho_0, $ and $K_0$ for SNM, $\tilde\alpha$
is calculated as:
\bea
\label{xeq36}
\tilde{\alpha}=\frac{\frac{K_0} {9} + \frac{E_F^0}{3} \left(\frac{12}{5}-2 
\frac{m}{m_0^{\star}}\right)}{\frac{E_{F}^{0}}{5}\left
(3-2 \frac{m}{m_0^{\star}}\right)-e_{0}}.
\eea

\subsection{The nuclear matter incompressibility}

In Ref. \cite{Blaizot1980} a linear correlation between $K_0$ and the ISGMR energy 
$E_{GMR} $ for a heavy
nucleus like $^{208}$Pb calculated with various Skyrme  EDFs was shown.
This correlation led to $K_0=  210\pm 30$ MeV when subjected to the
experimental value of ISGMR energy of the said nucleus.
After several revisions from
different corners a near settled value of $K_0=230 \pm 20$ MeV was posited
\cite{ToddRutel2005,Niksic2008,Avogadro2013}.  The recent ISGMR data
on Sn and Cd isotopes were, however, found to be
incompatible with this value of $K_0$ .
        These nuclei showed remarkable softness towards compression,
        apparently the ISGMR data appeared best explained with
        $K_0\sim200$ MeV
\cite{Avogadro2013}.

\begin{figure}
\centering
\resizebox{0.75\columnwidth}{!}{%
\includegraphics{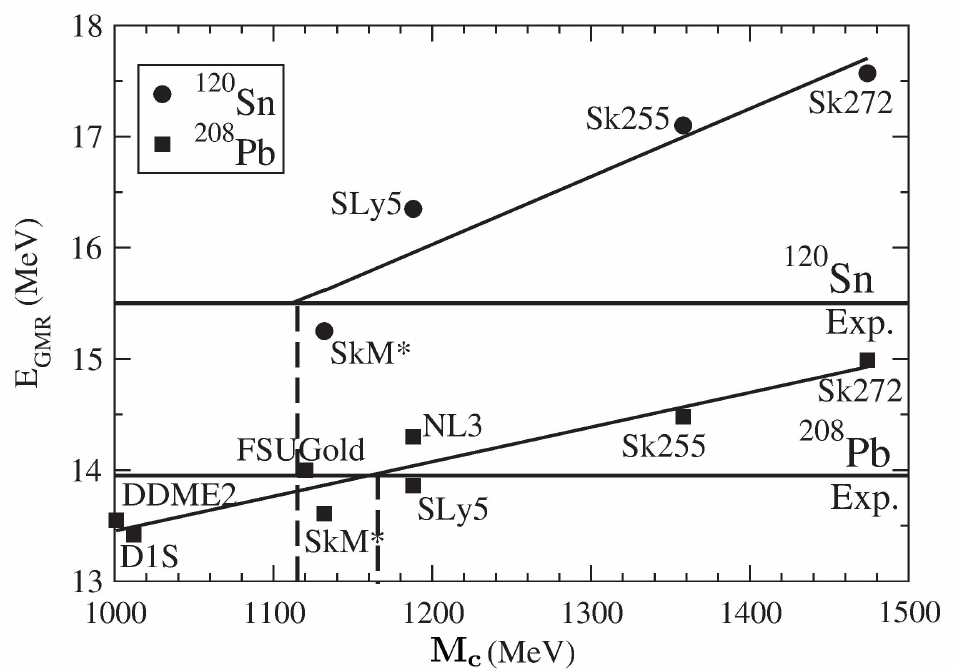} }
\caption{Centroid of the ISGMR in $^{208}$Pb and $^{120}$Sn calculated
with the constrained Hartree-Fock  method versus  the value of $M_c$
for various functionals. The experimental values for $^{208}$Pb and
$^{120}$Sn are taken from Refs.\cite{Lui2004,Li2007,Youngblood2004},
respectively, with respective error bars of $\pm200$ and $\pm 100$
keV. This figure is taken from Ref.\cite{Khan2012}.}
\label{fig:1} 
\end{figure}

        Rigorous analysis questioned the validity of the assumption of
        a strong correlation between $K_0$
and $E_{\scriptsize{GMR}}$ \cite{Colo2004,Khan2013} calculated from
        different forces, arguments were placed in favour of the
        fact that the ISGMR centroid $E_{\scriptsize{GMR}}$ maps the integral of the
        incompressibility $\int K(\rho) d\rho$ over the whole density
        rather than a single value at $\rho_0$.  A larger value of
        $K(\rho_0)$ for a given EDF can be compensated by a lower value
        of $K(\rho)$ at sub-saturation density so as to predict a similar
        value of ISGMR energy; as a result, $E_{\scriptsize{GMR}}$ might be a reflection
        of nuclear matter  incompressibility at an effective density lower than
        $\rho_0$. Indeed, it is seen that $K(\rho)$ calculated with
        a multitude of EDFs when plotted against density cross each
        other at a density  $\rho_c$ $[=(0.710 \pm 0.005) \rho_0]$
        \cite{Khan2012}. This  universality possibly arises from
        the constraints encoded in the EDF from empirical nuclear
        observables. This crossing density $\rho_c$ looks  more
        relevant as an indicator for the ISGMR centroid, with $K_c(\equiv
        K(\rho_c)) $ found to be around $35 \pm 4$ MeV \cite{Khan2013}. As
        the centroid energy $E_{\scriptsize{GMR}}$ maps the
        incompressibility integral, it seems, $E_{\scriptsize{GMR}}$
        is more intimately correlated with $M(\rho_c)$, the density
        derivative of $K(\rho)$ at the crossing density. The calculated
        values of $M(\rho_c)$ from various EDFs are found to be linearly
        correlated with the corresponding $E_{\scriptsize{GMR}}$'s
        for both $^{208}$Pb and $^{120}$Sn (see Fig. \ref{fig:1}).
        From known experimental ISGMR data for the nuclei, a value of $M_c
        (\equiv M(\rho_c)) \approx 1050 \pm 100$ MeV \cite{Khan2013} is then
        obtained.  Coming back to Eq. (\ref{xeq14}), with known values of
        $\rho_c$, $K_c$ and $M_c$ in conjunction with Eq. (\ref{xeq15}),
        the value of $a$ and $\tilde \alpha$ are calculated (we already
        assumed $e_0$, $\rho_0$ and $k_+$ to be known). The value of $K_0$
        can now be related as

        \bea \label{xeq17}
        K\left(\rho_{0}\right)=K\left(\rho_{c}\right)+&\left(\rho
        -\rho_{c}\right)
        K^{\prime}\left(\rho_{c}\right)+\frac{\left(\rho-\rho_{c}\right)^{2}}{2}
        K^{\prime \prime}\left(\rho_{c}\right)+ \nonumber \\
&+\left(\frac{\rho-\rho_{c}}{6}\right)^{3} K^{\prime \prime
\prime}\left(\rho_{c}\right)+\cdots
         \eea
        The higher derivatives of $K(\rho)$ can be calculated recursively
        from Eq. (\ref{xeq14}).  With chosen values of $e_0 \sim -16.0$
        MeV, $\rho_0\sim0.155$ fm$^{-3}$ and $m^{\star}/m \sim 0.7$,
        $K_0$ comes out to be $\sim 212$ MeV \cite{De2015}.
The first term in Eq. (\ref{xeq17}) is 35 MeV,
        the second term turns out to be 143.3 MeV, the third term is
        $35.9$ MeV, the fourth term is $-3.2$MeV and so on which adds
        up to $\sim 212$ MeV. Since `$a$' and $\tilde \alpha$ are known,
        $Q_0$ and the other higher
derivatives entering in Eq. (\ref{xeq3}) can be
        calculated thus defining the nuclear matter  EoS, the first few terms
        giving a nearly precise description of the EoS around the
        saturation density. The value of $Q_0$ turns out to be $\sim -378 $ MeV. 

          \section{Asymmetric nuclear matter}

We now consider asymmetric nuclear matter (ANM), a two component system consisting of neutrons and protons. 
          To lowest order in the asymmetry parameter $\delta$, the energy
          of ANM contains, in addition to
          the SNM term (see Eq. (\ref{xeq1})) a
contribution $C_2(\rho) \delta^2$. The knowledge of
          $C_2(\rho)$ Eq. (\ref{xeq4}) involves understanding of 
          its higher order density derivatives like $L_0$, $K_{\rm sym}^0$,
          etc. around $\rho_0$, in addition to $C_2^0$, the
          symmetry energy coefficient of ANM at $ \rho_0$.
In the following, we discuss on constraining  the values of the symmetry
elements through their correlations with various observables pertaining to
finite nuclei and neutron stars. We further explore the possibility of
plausible correlations among the symmetry elements themselves to see how
the knowledge of known lower order symmetry elements can throw light on
the values of the unknown higher order symmetry elements.

        \subsection{The symmetry energy coefficient}

          The value of  $C_2^0$ is now known 
           in very tight bounds \cite{Jiang2012}. The higher order density derivatives are, however, not
very precisely known. 
The microscopic-macroscopic mass formula improved 
          with the consideration of mirror nuclei 
constraint \cite{Wang2010} describes 
          the binding energies in the entire nuclear mass table extremely well; 
          removing the contributions of volume, surface, Coulomb, pairing and
          Wigner terms of this mass formula from the experimental binding energies, 
          one is left with the symmetry energy that is called the `experimental 
          symmetry energy' $S(Z,A)$ for a nucleus with charge $Z$ and mass $A$. 
It may be a little approximate representation 
          of the symmetry energy because of  
remnants of the effects due to shell 
          structure, but the double differences of these experimental symmetry energies 
          of neighboring nuclei (denoted by $\mathcal {R}_{ip-jn}(Z,A)$) effectively 
          cancel the shell effects and then the double difference becomes a powerful
          tool for extracting the symmetry energy elements from the observed compact 
          correlation of $\mathcal{R}_{ip-jn}(Z,A) [\mathcal{R}_{ip-jn}]$ with mass number 
          $A$. The double difference
          is defined as

          \bea
          \label{xeq18}
          \mathcal{R}_{i p-j n}(Z, A)=&S(Z, A)+S(Z-i, A-i-j) \nonumber \\
           &-S(Z, A-j)-S(Z-i, A-i).
          \eea
\begin{figure}
\centering
\resizebox{0.75\columnwidth}{!}{%
\includegraphics{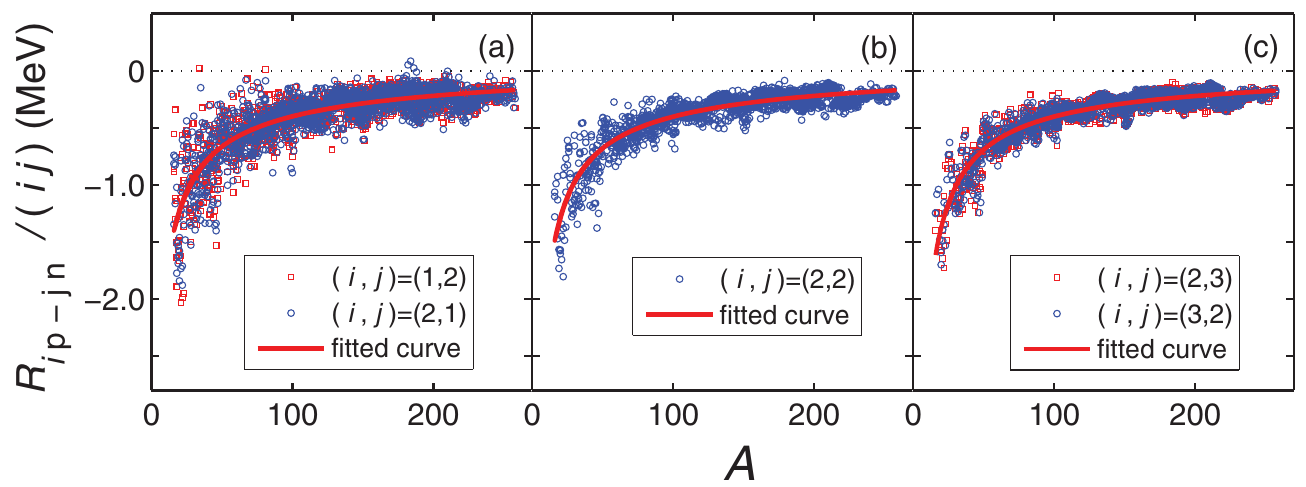} }
\caption{$R_{{ip-jn }} /(i j)$ (MeV) versus mass number $A$ . $R_{
{ip }-{j n}}$ is given by double  difference of experimental symmetry
energies of four neighboring nuclei. Refer to the text for details. Panel
(a): $(i, j)=(1,2)$ and (2,1),  (b)$ (i, j)$ =(2,2)  and (c)$(i,
j)=(2,3)$ and $(3,2)$ The solid red lines are used to guide the eyes. The
figure is taken from Ref.  \cite{Jiang2012}.} \label{fig:2} 
\end{figure}

          The values of $\mathcal{R}_{ip-jn}(Z,A)/(i j)$ are displayed in 
          Fig. \ref{fig:2}  for  $(i,j)=(1,2),(2,1), (2,2)$ and $(3,2)$. 
         No noticeable 
          difference is seen between even-odd and odd-even nuclei, or between 
          even-even and odd-odd nuclei for 
          $\mathcal{R}_{ip-jn}$. Assuming 
          \bea
          \label{xeq19}
          S(Z,A)= a_{\rm sym}(A) I^2 A = (C_{\rm sym}^v - 
          C_{\rm sym}^s A^{-1/3}) I^2 A ,
          \eea       
          an expression for  $\mathcal{R}_{ip-jn}$ 
          can be obtained in $C_{\rm sym}^v$ and $C_{\rm sym}^s$, which 
          when fitted across the mass spectrum yields value for 
           $C_{\rm sym}^v=32.10 \pm 0.31$ MeV and for $C_{\rm sym}^s=58.91 
          \pm 1.08$ MeV. In Eq. (\ref{xeq19}), $a_{\rm sym}(A)$ 
is the symmetry energy 
 coefficient of a finite nucleus, 
the volume symmetry energy $C_{\rm sym}^v$ is 
          identified with $C_2(\rho_0)$ ($\equiv C_2^0$), 
          $C_{\rm sym}^s$ is the surface symmetry energy 
          coefficient and $I=(N-Z)/A$, the equivalent to the asymmetry parameter 
          $\delta $ of ANM. The volume symmetry energy, so obtained does 
          not differ significantly from that ($C_2(\rho_0)=31.95 \pm 1.75$MeV) 
          obtained from analysis of excitation to 
isobaric analog states \cite{Danielewicz2014} 
          augmented with empirical values of neutron skins determined using hadronic probes. 
          
         \subsection{The density derivatives of symmetry energy}
\begin{figure}
\centering
\resizebox{0.75\columnwidth}{!}{%
\includegraphics{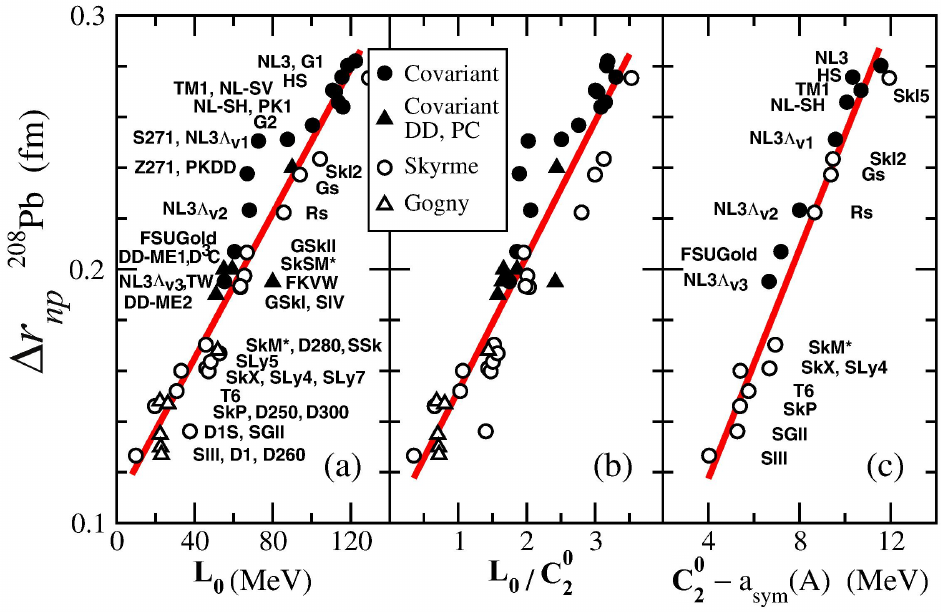} }
\caption{Correlation of the  neutron-skin thickness $\Delta r_{np}$  for 
$^{208}$Pb with the slope of the symmetry energy $L_0$ (a), the ratio $L_0/C_2^0 $
(b), and with $C_2^0 -a_{\rm sym}(A)$ (c), for various nuclear models ( DD
and PC stand for density dependent and point coupling models). From left
to right, the correlation factors are $r=0.961,0.945,$ and $0.970$. The
figure is taken from Ref.\cite{Centelles2009}. }

\label{fig:3}      
\end{figure}

          The density derivatives of the symmetry 
energy coefficients ($L_0,K_{\rm sym}^0$ etc)
          are not yet known with desirable   certainty. Calculations with
selective Skyrme  and relativistic 
          EDFs show a nearly linear correlation of $C_2^0$
          with $L_0$  \cite{Carbone2010,Tsang2012,Danielewicz2014,Lattimer2016} 
pointing to a value of the symmetry derivative $L_0$. 
          In a considerable density range around $\rho_0$, it has been found that the ansatz 
          \bea
          \label{xeq20}
          C_2(\rho)= C_2 (\rho_0) (\frac{\rho}{\rho_0})^\beta,  
          \eea

          works well \cite{Chen2005,Li2008}, 
          where $\beta$ is a constant $\sim 0.69$. Then 
          $L_{0}=\left.3 \rho_0 \frac{\partial C_{2}} {\partial \rho}\right|_{\rho_{0}}=3 \beta C_{2}^{0}.
          $ In Ref. \cite{Carbone2010}, with a few Skyrme interactions.
a nearly linear correlation of $L_0$ with $C_2^0$ has been reported. 
           This is, however, approximate. The tendency of a larger
          $L_0$ with larger $C_2^0$ can not be overlooked
          though. Different nuclear physics observables like isospin
          diffusion, nuclear emission ratio, isoscaling,  giant dipole
          resonances, pygmy dipole resonance etc 
\cite{Li2008,Famiano2006,Shetty2007,Trippa2008,Carbone2010} hint at a central value
          of $L_0$, all differing from each other, astronomical data on
          neutron star masses and radii providing a further different
          value \cite{Steiner2012}. 
In a novel exercise involving the nuclear Droplet
          Model (DM), Centelles {\it et al} \cite{Centelles2009,Warda2009}
          showed that the neutron-skin $\Delta r_{np}$ of a nucleus
          can be recast to leading order in $L_0$ lending a good linear
          fit of $\Delta r_{np}$ with $L_0$ (see Fig. \ref{fig:3}). With
          $\Delta r_{np}$ known from hadronic probes, a value of $L_0=
          75\pm 25$ MeV was arrived at. In the ambit of microscopic
          calculations with different EDFs, this idea was advanced
          further, the correlation of the neutron-skin thickness with $L_0$ helped
          to find $L_0$ in narrower limits \cite{Agrawal2012,Agrawal2013}.
{\bl However the estimates on neutron skin thickness based on the
hadronic probes are model dependent. The $^{208}$Pb Radius EXperiment (PREX)
and $ ^{48}$Ca Radius EXperiment CREX experiments based on the electroweak
probe would allow the model independent determination of the neutron
skin thickness in $^{208}$Pb and $^{48}$Ca nuclei. These experiments are
designed to extract the neutron skin thickness from parity violating
electron scattering. The extracted $^{208}$Pb skin thickness $\Delta
r_{np}=0.33^{+0.16}_{-0.18}$ fm \cite{Abrahamyan2012} has very  large
statistical uncertainty. The future experiment PREX-II is designed to
achieve the originally proposed experimental precision in $\Delta r_{np}$
to $1$ \% \cite{PREX-II}.  The CREX is expected to provide precision
in $\Delta r_{np}$ of $^{48}$Ca to $0.6$\% \cite{CREX}. The $^{48}$Ca
being a light nucleus  may  provide the key information for bridging ab
initio calculations and those based on the density functional theories. }

The shroud of uncertainty looms  larger on the higher symmetry
derivatives $K_{\rm sym}^0 (=K_{\rm sym}(\rho_0)=9\rho_0^2
\frac{\partial^2 C_2}{\partial \rho^2}|_{\rho_0}$)  and $Q_{\rm
sym}^0(=Q_{\rm sym}(\rho_0)=27 \rho_0^3 \frac{\partial^3 C_2}{\partial
\rho^3}|_{\rho_0})$. The values of $K_{\rm sym}^0$ and $Q_{\rm sym}^0$,
in different parameterizations of the Skyrme EDFs lie in very wide
ranges [-700 MeV $< K_{\rm sym}^0 < +400 $MeV, -800 MeV $< Q_{\rm
sym}^0 < 1500$ MeV] \cite{Dutra2012,Dutra2014}. From the ansatz in
Eq. (\ref{xeq20}), it is seen that $K_{\rm sym}^0 = 3(\beta - 1)L_0$ and
so should have a perfect linear correlation with $L_0$ when $\beta$
is a constant. Danielewicz and Lee \cite{Danielewicz2009} have shown a
linear relationship between $K_{\rm sym}^0$ and $L_0$, studied with 118
Skyrme EDFs,. Tews {\it et al} \cite{Tews2017} propose a relation of the
form: 
\bea
\label{xeq21}
K_{\rm sym}^0 \approx 3.41 L_0 - 306 \pm 40 ~{\rm MeV}. 
\eea 

{\bl Mondal {\it et al} \cite{Mondal2018}, with a total of 237 Skyrme EDFs also
find a correlation between them, but the correlation is not very strong
(correlation coefficient $r\sim 0.8$); the correlation becomes more robust
only when a selected subset of 162 EDFs are chosen  ($r\sim 0.91$).
These 162  Skyrme EDFs are selected by constraining the iso-scalar nucleon
effective mass $\frac{m_0^*}{m} $ to $0.85 \pm 0.15 $ and the isovector
splitting of effective mass $\mid{\frac{m_n^\star-m_p^\star}{m}}\mid$
to less than unity, which more than covers the values from the limited
experimental data \cite{Zhang2016,Coupland2016,Kong2017} and recent
theoretical values on it.}

      \subsection{Interrelating symmetry elements}

 Starting from a plausible set of approximations on the  nucleonic
interaction (density dependent, quadratically dependent on momentum)
as  stated in the beginning of Sec.3, for asymmetric nuclear
matter, the  equation for the energy per nucleon can be generalized from
Eq. (\ref{xeq11}) to 
\bea
\label{xeq22}
e(\rho, \delta)=\frac{1}{\rho}\left[\sum_{\tau} 
\frac{P_{F, \tau}^{2}}{10 m} \rho_{\tau}\left(3-2 
\frac{m}{m_{\tau}^{ \star }(\rho)}\right)\right]-V_{2}
(\rho, \delta)+\frac{P(\rho, \delta)}{\rho}.
\eea
Here $\tau$ is the isospin index, $\rho_\tau= (1+\tau\delta)\frac{\rho}{2}; \tau = 1$ for
neutrons and -1 for protons. The Fermi momentum of the component nucleonic matter can be written
as $P_{F,\tau}=g_2\rho_\tau^{1/3}$ with $g_2 = (3\pi^2)^{1/3}\hbar$. The effective mass for the
two component nuclear matter, to lowest order in $\rho$ is written as 
\bea
\label{xeq23}
\frac{m}{m_{\tau}^{\star}(\rho)}=1+\frac{k_{+}}{2} \rho+\frac{k_{-}}{2} 
\rho \tau \delta,
\eea 
and the rearrangement potential generalized for ANM 
\bea
\label{xeq24}
V_{2}(\rho, \delta)=\left(a+b ~\delta^{2}\right) \rho^{\tilde \alpha}.
\eea
The new constant $b$ is a measure of the asymmetry dependence of the rearrangement potential.
Since, $P=\rho^2\frac{\partial e}{\partial \rho}$, 
Eq. (\ref{xeq22}) can be integrated
\cite{Malik2018a}
\bea
\label{xeq25}
e(\rho, \delta)=& \frac{3}{10} \gamma\left[\sum_{\tau}\left
(1+\tau ~\delta\right)^{5 / 3}\left\{\rho^{2 / 3}+\frac{1}{2} 
\rho^{5/3}\left(k_{+}+k_{-} \tau ~ \delta\right)\right\}\right] \nonumber \\
&+\left(a+b ~\delta^{2}\right) \frac{\rho^{\tilde \alpha}}
{(\tilde \alpha-1)}+K(\delta) \rho,
\eea
where, $\gamma = g_2^2/(2^{5/3}m)$ and $K(\delta)=(K_1 + K_2\delta^2 + K_4\delta^4+...)$ is a
constant of integration. The symmetry coefficient $C_2(\rho)
(=\frac{1}{2}\frac{\partial^2e}{\partial \delta^2}|_{\delta = 0})$ is then derived as 
\bea
\label{xeq26}
C_{2}(\rho)=\frac{b \rho^{\tilde \alpha}}{(\tilde \alpha-1)}+
\frac{\gamma}{3} \rho^{2 / 3}\left[1+\frac{1}{2}\left(k_{+}
+3 k_{-}\right) \rho\right]+K_{2} \rho,
\eea
so also $L(\rho)$ and $K_{\rm sym}(\rho)$, 
\bea
\label{xeq27}
L(\rho)=\frac{3 \tilde \alpha}{(\tilde \alpha-1)} 
b \rho^{\tilde \alpha}+\frac{2}{3} \gamma \rho^{2 / 3}+
\frac{5}{6} \gamma \rho^{5 / 3}\left(k_{+}+3 k_{-}\right)
+3 K_{2} \rho,
\eea

\bea
\label{xeq28}
K_{\rm sym}(\rho)=9 \tilde \alpha b {\rho^{\tilde\alpha}}-\frac{2}{3} 
\gamma \rho^{2 / 3}+\frac{5}{3} \gamma \rho^{5 / 3}\left(k_{+}+3 k_{-}\right).
\eea
A little algebra then leads to an equation interrelating $C_2^0, L_0$ and $K_{\rm sym}^0$
\bea
\label{xeq29}
K_{\rm sym}^{0}=-5\left[3 C_{2}^{0}-L_{0}\right]+
3 b \rho_{0}^{\tilde{\alpha}}(3 \tilde{\alpha}-5)+E_{F}^{0}.
\eea
We note at this point that the r.h.s of Eq. (\ref{xeq22}) 
can be expanded in powers of $\delta$
using expressions for $P$ and $V_2(\rho,\delta)$. Comparing with 
Eq. (\ref{xeq1}) and equating
coefficients of the same order in $\delta$ one gets 
an expression for $C_2(\rho)$ (it looks
somewhat different \cite{Mondal2017} from that given in Eq. (\ref{xeq26})
though they are equivalent)
\bea
\label{xeq30}
C_{2}(\rho)=-b \rho^{\tilde \alpha}+\rho \frac{\partial C_{2}}
{\partial \rho}+\frac{\gamma}{9} \rho^{2 / 3}\left[1-\left(k_{+}
+3 k_{-}\right) \rho \right].
\eea
This leads to expressions for $K_{\rm sym}^0$ and $Q_{\rm sym}^0$ as, 
\bea
\label{xeq31}
K_{\rm sym}^{0}=-3 \tilde{\alpha}\left[3 C_{2}^{0}-
L_{0}\right]+E_{F}^{0}\left[(3 \tilde{\alpha}-4)+\left(\frac{2}{3}  
\frac{m}{m_{0}^{\star}}+k_{-}~ \rho_0\right)(5-3 \tilde{\alpha})\right],
\eea
and 
\bea
\label{xeq32}
Q_{\rm sym}^{0}=15 \tilde{\alpha}\left[3 C_{2}^{0}-
L_{0}\right]+K_{\rm sym}^{0}(3 \tilde{\alpha}-1)+E_{F}^{0}(2-3 \tilde{\alpha}).
\eea
Eqs. (\ref{xeq31}) and (\ref{xeq32}) show that   there is an 
interrelationship between $K_{\rm sym}^0$ and $Q_{\rm
sym}^0$. Making use of Eq. (\ref{xeq29}), one gets :
\bea
\label{xeq33}
K_{\rm sym}^{0}+Q_{\rm sym}^{0}=
9 b \rho_{0}^{\tilde{\alpha}}  \tilde{\alpha}(3 \tilde{\alpha}-5)
+E_{F}^{0}.
\eea
Eq. (\ref{xeq29}) looks very similar to that obtained for the 
correlations among $C_2^0$, $ L_0$ and $K_{\rm sym}^0$ in 
Skyrme models \cite{Mondal2018}.
This is not coincidental. There is an exact equivalence of the Skyrme functional with the EDF
given by Eq. (\ref{xeq25}) provided the term $K(\delta)$ is truncated at $\delta^2$. The parameters
$\tilde{\alpha }, K_1, K_2$ etc. can then be correlated to the standard Skyrme parameters
\cite{Malik2018a}
\bea
\label{xeq34}
&&\tilde \alpha =\alpha+1 \nonumber \\
&&K_{1} =\frac{3}{8} t_{0} \nonumber \\
&&K_{2} =-\frac{1}{4} t_{0}\left(x_{0}+\frac{1}{2}\right) \nonumber \\
&&a =\frac{1}{16} t_{3} \alpha \nonumber \\
&&b =-\frac{1}{24} t_{3}\left(x_{3}+\frac{1}{2}\right) \alpha  \nonumber \\
&&k_{+} =\frac{m}{\hbar^2}\left[\frac{3}{4} t_{1}+\frac{5}{4} t_{2}+t_{2} x_{2}\right] \nonumber \\
&&k_{-} =\frac{m}{2 \hbar^{2}}\left[t_{2}\left(x_{2}
+\frac{1}{2}\right)-t_{1}\left(x_{1}+\frac{1}{2}\right)\right].
\eea

\begin{figure}
\centering
\resizebox{0.75\columnwidth}{!}{%
\includegraphics{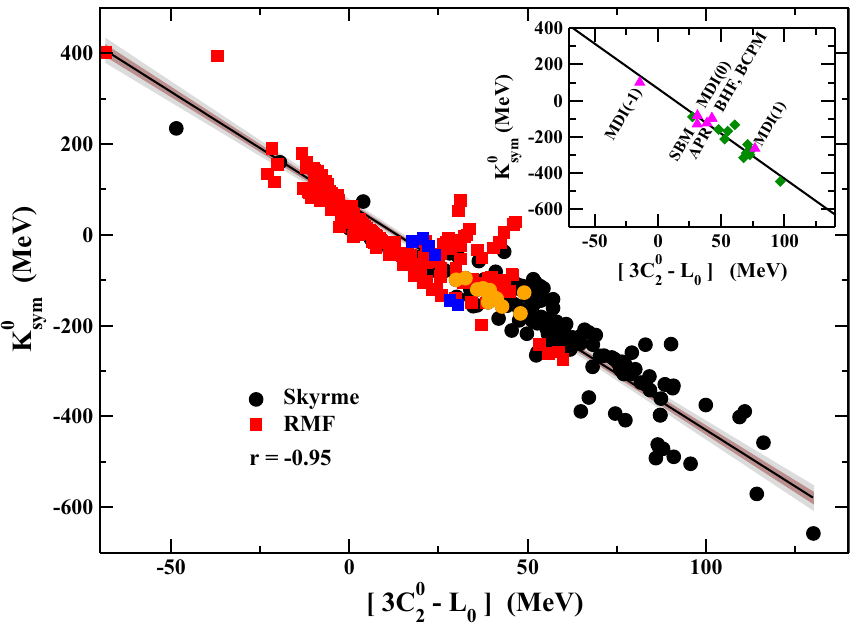} }
\caption{The correlation between $K_{\mathrm{\rm sym}}^{0}$ and 
$\left[3 C_{2}^{0}-L_{0}\right]$ as obtained from 500 EDFs
\cite{Dutra2012,Dutra2014}.  The black circles correspond to the
Skyrme-inspired EDFs, and the red squares refer to those obtained from
RMF models.  The models consistent with all the constraints demanded
by Dutra et al.\cite{Dutra2012} are highlighted by orange circles
for Skyrme EDFs \cite{Dutra2012} and blue squares for RMF EDFs
\cite{Dutra2014}.  The inner (outer) colored regions around the
best-fit straight line through these points depict the loci of $95 \%$
confidence (prediction) bands of the regression analysis. The inset shows
the correlation line obtained from the Skyrme-RMF models, and the magenta
triangles are the results obtained from EDFs with realistic interactions,
$\mathrm{MDI}(0), \mathrm{MDI}(1), \mathrm{MDI}(-1)$\cite{Chen2009}
APR \cite{Akmal1998}, BHF \cite{Taranto2013}, BCPM \cite{Baldo2013},
and $\mathrm{SBM}$\cite{Agrawal2017}, respectively.  The green diamonds
represent results from a few Gogny interactions \cite{Sellahewa2014}. The
figure is taken from \cite{Mondal2017}.}
\label{fig:4}       
\end{figure}

\begin{figure}
\centering
\resizebox{0.75\columnwidth}{!}{%
\includegraphics{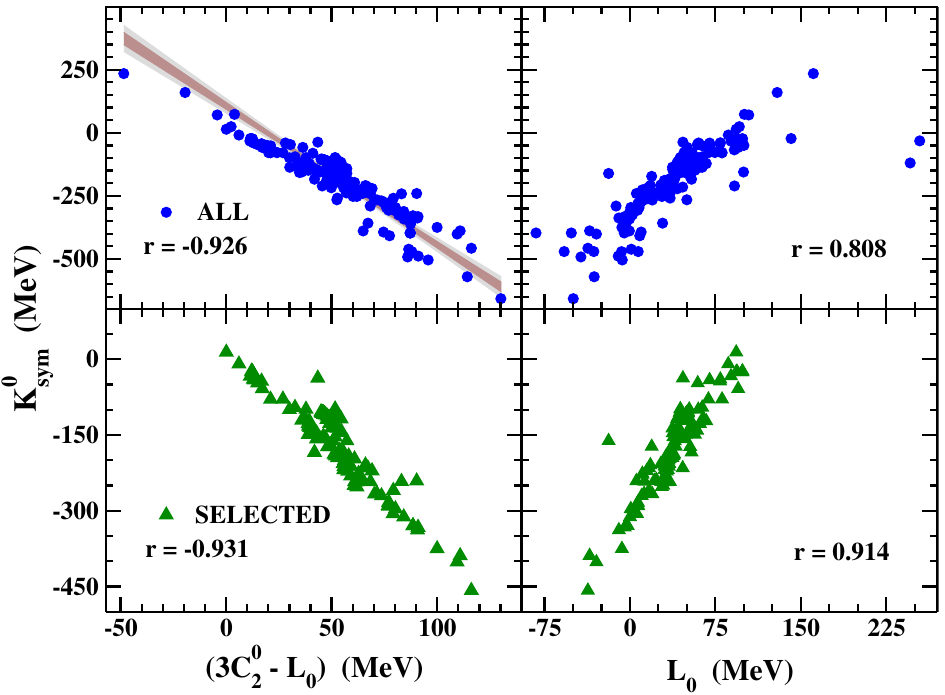} }
\caption{The correlation of $K_{\rm sym}^0$ with $L_0$ and 
with $\left[3 C_{2}^{0}-L_{0}\right]$ are depicted 
in the right and left panels
 respectively. Results for 237 Skyrme EDFs ('ALL') 
are displayed in the upper panels, the lower panels 
contain the results for a selected subset of 162 models 
('SELECTED') (see text for details). 
The inner(outer) colored regions around the best-fit straight 
line in upper left panel depict the loci 
of $95\%$ confidence (prediction) bands of the regression analysis. The
figure is taken from Ref. \cite{Mondal2018}.}

\label{fig:5}       
\end{figure}

\begin{figure}
\centering
\resizebox{0.75\columnwidth}{!}{%
\includegraphics{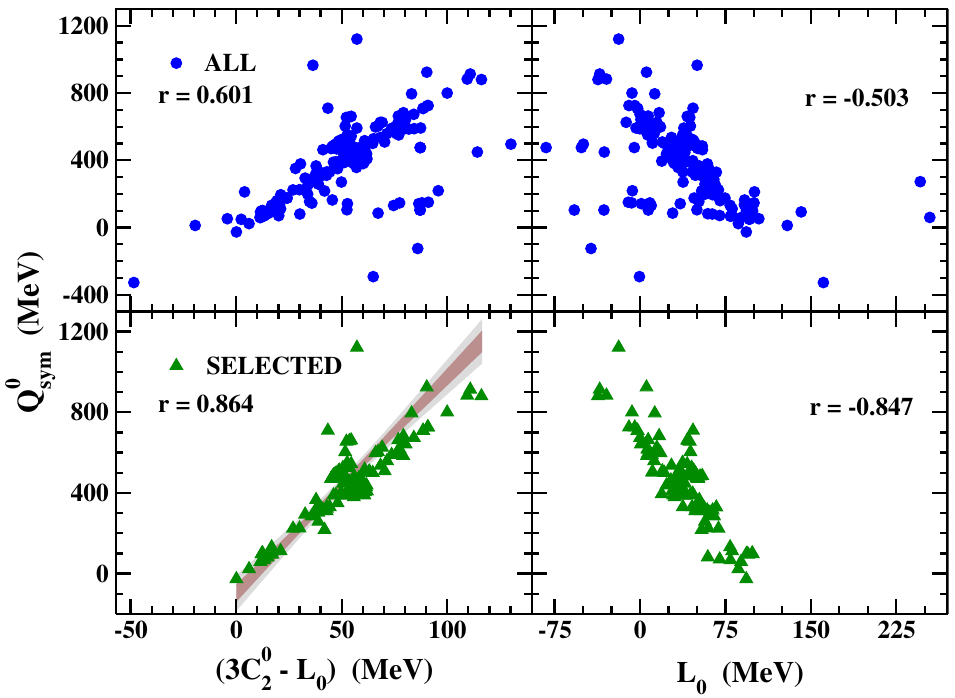} }
\caption{Same as Fig.\ref{fig:5}, but for $Q_{\rm sym}^0$. 
The confidence bands of regression analysis are 
given only for the subset of models ('SELECTED') in 
the lower left panel. The figure is taken from Ref.\cite{Mondal2018}.}
\label{fig:6}       
\end{figure}

The structure of Eq. (\ref{xeq29})  and Eq. (\ref{xeq31}) shows that there
is a strong likelihood of a linear correlation  between $(3C_2^0 - L_0)$
and $K_{\rm sym}^0$. Realizing that the EDF was obtained from general
thermodynamical considerations and some very plausible assumptions on
the nature of the nuclear force, one may expect this correlation to be
universal; this is vindicated from the correlated structure of $K_{\rm
sym}^0$ with $(3C_2^0 - L_0)$ as displayed in Fig. \ref{fig:4}
for 500 EDFs \cite{Dutra2012,Dutra2014} that have been in use to
explain nuclear properties.  The correlation is seen to be very robust,
the correlation coefficient $r = -0.95$ . In the inset of the figure,
results corresponding to EDFs obtained from several realistic interactions
(magenta triangles) and a few Gogny interactions (green diamonds) are also
displayed. They lie nearly on the correlation line highlighting further
the universality in the correlation.  Imposing a general constraint
that the neutron energy per particle should be zero at zero density of
neutron matter, a plausible explanation of such a correlation was given
recently \cite{Margueron2018}. The linear regression analysis yields

\bea
\label{xeq35}
K_{\rm sym}^{0}=d_{1}\left(3 C_{2} - L_{0}\right)+d_{2},
\eea
with $d_1 = -4.97\pm 0.07$ and $d_2 = 66.80\pm 2.14$ MeV.  One sees
that $d_1$ is very close to $-5$ as expected form Eq. (\ref{xeq29}). In
a recently developed density dependent Van der Waals model for nuclear
matter, with some constraints on $K_0$, exactly such a relation was found
\cite{Dutra2020} where $d_1=-6.3$ and $d_2=51.5$ MeV.  As mentioned
earlier, from Eq. (\ref{xeq21}) one also expects a correlation
between $K_{\rm sym}^0$ and $L_0$, but it is comparably weaker (see
Fig. \ref{fig:5}). With a total of 237 Skyrme EDFs Mondal {\it et al}
\cite{Mondal2018} looked for correlation between $Q_{\rm sym}^0$ and
$(3C_2^0 - L_0)$ and also with $L_0$. The correlation found was poor,
only with a selected subset of Skyrme models  as discussed earlier, 
an improved correlation was found (see Fig. \ref{fig:6} ).

Estimates of the symmetry elements $L_0, K_{\rm sym}^0$ etc., were made
by Centelles {\it et al} \cite{Centelles2009} from the correlation
systematics of $L_0$ with the neutron-skin of nuclei obtained
from hadronic probes. The neutron skins have large uncertainties,
so do $L_0$. From equations we have set up, we show that knowledge
of the symmetry energy at another density (but for $\rho_0$), say
a sub-saturation density $\rho_1$, helps to find a more controlled
value of $L_0$.  We choose the value of $C_2(\rho_1) (=24.1\pm 0.8 $
MeV) at $\rho_1 = 0.1$ fm$^{-3}$ found from its strong correlation
with the centroid of the GDR resonance energy in spherical nuclei
in Skyrme EDFs \cite{Trippa2008}. An initial estimate of $L_0$ can
be done from Eqs. (\ref{xeq4}) and (\ref{xeq35}).  Leaving out terms
beyond $\epsilon^2$ in Eq. (\ref{xeq4}) (which may not be a very bad
approximation), with $C_2^0 = 32.1\pm 0.3$ MeV, $\rho_0 = 0.155\pm 0.008$
 fm$^{-3}$, $C_2(\rho_1)$ and $\rho_1$ being just
mentioned, one gets $L_0 \sim 61.3$ MeV. A more dependable value is,
however, obtained from an input value of the effective mass  $m_0^*$.

Then, keeping terms upto  $Q_{\rm sym}^0$ in Eq. (\ref{xeq4}), 
from Eqs. (\ref{xeq32}) and (\ref{xeq35}),
$L_0$, $K_{\rm sym}^0$, $Q_{\rm sym}^0$ can be calculated. For the effective mass, a value
of $m_0^*/m = 0.70\pm 0.05$ is taken that is consistent with many analyses
\cite{Jaminon1989,Li2015}. The value of the symmetry elements then turn out to be $L_0=
60.3\pm 14.5$ MeV, $K_{\rm sym}^0 = -111.8\pm 71.3$ MeV, and $Q_{\rm sym}^0 = 296.8\pm
73.6$ MeV. The value of $k_-$ can be calculated from Eq.(\ref{xeq31}). 
It is a measure of
the isovector effective mass splitting 
$\Delta m_0^*$ at asymmetry $\delta$, 
\bea
\label{xeq37}
\Delta m_{0}^{\star}=\left(\frac{m_{n}^{\star}
-m_{p}^{\star}}{m}\right)_{\rho_{0}} \cong-k_{-} 
\rho_{0}\left(\frac{m_{0}^{\star}}{m}\right)^2 \delta.
\eea
Its value turns out to be $\Delta m_0^* = (0.17\pm 0.24)\delta$.

\subsection{Electric dipole polarizability: relations to symmetry elements}

Under the action of an isovector probe, semi-classically speaking, the
centers of the neutron and the proton fluids separate leading to an
electric dipole polarization in the nucleus. The dipole polarizability
$\alpha_D$ is defined as,
\bea
\label{xeq38}
\alpha_{D}=\frac{8 \pi e^{2}}{9} \int_{0}^{\infty} 
\omega^{-1} \mathcal{R}\left(\omega, E_{1}\right) 
d \omega=\frac{8 \pi e^{2}}{9} m_{-1}(E_1).
\eea

Here $R(\omega,E1)$ is the electric dipole strength as a function of the
excitation energy $\omega$ and $m_{-1} (E1)$ is called the inverse energy
weighted sum rule for the electric dipole $(E1)$ excitations. The dipole
polarizability is an experimentally measurable quantity and thus becomes
an effective indicator of the symmetry elements related to asymmetric
nuclear matter. The $m_{-1}$ moment may be obtained with the random-phase-approximation ( RPA) 
methodology, the so-called dielectric theorem \cite{Bohigas1979,Hinohara2015}
also allows to extract it from a constrained ground state calculation. It
is related to the constrained energy $E_x$ \cite{Bohigas1979} as 
\bea
\label{xeq39}
\left(\frac{m_{-1}}{m_{1}}\right)^{1 / 2}=1 / E_{x},
\eea
where $m_k$ is the $kth$ moment of the energy weighted sum rule (EWSR) 
\bea
\label{xeq40}
m_{k}=\int d \omega \quad \omega^{k} \quad \mathcal{R}\left(\omega, E_{1}\right).
\eea 
Solving the constrained problem classically in the ambit of the Droplet Model (DM)
\cite{Myers1974}, it was shown \cite{Meyer1982} that for a nucleus of mass $A$ , the dipole
polarizability can be written in terms of the symmetry energy constant $C_2^0$ as 
\bea
\label{xeq41}
\alpha_{D}^{\rm DM}=\frac{\pi e^{2}}{54} 
\frac{A\left\langle r^{2}\right\rangle}{C_2^0}\left(1+\frac{5}{3} \times 
\frac{9}{4} \frac{C_2^0}{Q} {A}^{- 1 / 3}\right).
\eea
In Eq.(\ref{xeq41}), $<r^2>$ is the mean-square radius of the nucleus and
$Q$ the surface stiffness constant, a measure of the resistance of the
motion of neutrons against protons.
Eq. (\ref{xeq41}) has a few consequences. The ratio $C_2^0/Q$ and the density
slope parameter $L_0$ are known to show a strong linear correlation
\cite{Warda2009,Centelles2009} for a large set of EDFs, the electric dipole
polarizability is therefore expected to be very closely tied to the
symmetry properties of asymmetric nuclear matter.  The DM relates the
symmetry energy coefficient $a_{\rm sym}(A)$ of a finite nucleus with
$C_2^0/Q$ as 
\bea
\label{xeq42}
a_{\rm sym}(A)=\frac{C_2^0}{1+\frac{9}{4} \frac{C_2^0}{Q}A^{-1 / 3}}.
\eea

Expanding Eq. (\ref{xeq42}) to first order in the small parameter $\frac{C_2^0}{Q}
A^{-1/3}$, Eq.  (\ref{xeq41}) may be written as 
\bea
\label{xeq43}
\alpha_{D}^{D M} \approx \frac{\pi e^{2}}{54} 
\times \frac{A <r^2>}{C_2^0} \quad\left(1+\frac{5}{3} 
\frac{C_2^0-a_{\rm sym}(A)}{C_2^0}\right).
\eea

\begin{figure}
\centering
\resizebox{0.75\columnwidth}{!}{%
\includegraphics{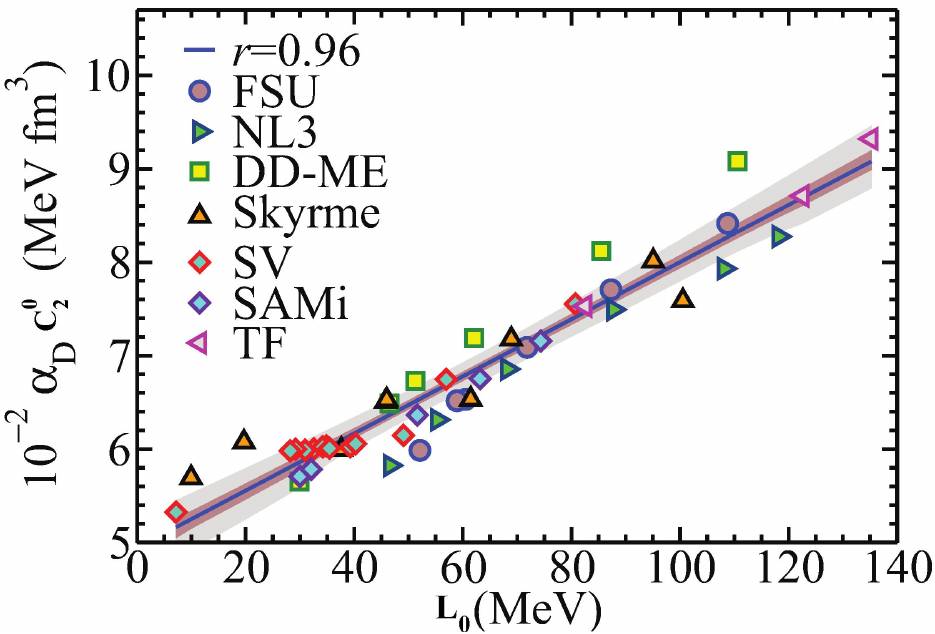} }
\caption{Dipole polarizability in $^{208}$ Pb times the symmetry energy
at saturation as a function of the slope parameter $L_0$ calculated with
some modern EDFs (see Ref.\cite{RocaMaza2013a} for further details).
The linear fit gives $10^{-2} \alpha_{D}
C_2^0=(4.80 \pm 0.04)+(0.033 \pm 0.001) L_0$ with a correlation coefficient
$r=0.96,$ and the two shaded regions represent the $99.9 \%$ and $70 \%$
confidence bands. This figure is taken from Ref.\cite{RocaMaza2013a} }.
\label{fig:7}       
\end{figure}

Since $C_2^0-a_{\rm sym}(A) = a_{\rm sym}^s A^{-1/3}$, where $a_{\rm sym}^s$ is the surface
symmetry term, it is evident that $\alpha_D^{DM}$ is sensitive to the ratio of the
surface and bulk symmetry energies. Since, 
\bea
\label{xeq44}
a_{\rm sym}(A) \approx C_{2}\left(\rho_{A}\right),
\eea
where $\rho_A$ is close to the average density of a
nucleus (which is necessarily lower than $\rho_0$), Eq. (\ref{xeq43})
shows that the symmetry energy at the sub-saturation density $\rho_A$ can
be gauzed \cite{RocaMaza2013} if the dipole polarizability is known. In
Ref. \cite{Centelles2009}, $\rho_A$ for $^{208}$Pb is taken as $\sim 0.1$
fm$^{-3}$. In Ref. \cite{Agrawal2012}, it is shown how to evaluate it in
a local density approximation. From Eq. (\ref{xeq4}), to lowest orders,
$C_2(\rho_A)$ can be written as $C_2(\rho_A) = C_2^0+L_0\epsilon_A$, using
this in Eq. (\ref{xeq41}) results in

\bea
\label{xeq45}
\alpha_{D}^{\rm DM} \approx \frac{\pi e^{2}}{54}
\frac{A <r^{2}>}{C_2^0}\left[1-\frac{5}{3} 
\frac{L_{0}}{C_2^0} \epsilon_{A}\right],
\eea
where $\epsilon_A=(\rho_A-\rho_0)/3\rho_0$.  This formula is
suggestive of a close relationship between $\alpha_D$ and the symmetry
elements $L_0$ and $C_2^0$.  Indeed, it has been seen with Skyrme and six
different families of systematically varied EDFs that $\alpha_D C_2^0$
has a nearly linear relationship with $L_0$ \cite{RocaMaza2013} as
seen in Fig. \ref{fig:7}. Specifically with an adopted value of $C_2^0=
(31\pm 2_{\rm est})$ MeV, it is found that

\bea
\label{xeq46}
L_0= 43 \pm 6_{exp} \pm 8_{theo} \pm 12_{est} ~{\rm MeV},
\eea
where 'est' refers to the uncertainties derived from different estimates on $C_2^0$.

The simplicity of the DM allows one to extract a relationship
between $\alpha_D$ and the neutron-skin thickness $\Delta r_{np}$. In
terms of the bulk nuclear matter properties, the neutron skin in DM can
be written as \cite{Centelles2009},
\bea
\label{xeq47}
\Delta r_{np}=\sqrt{\frac{3}{5}}\left [\frac{3r_0}{2}\frac{\frac{C_2^0}{Q}(I-I_{\scriptsize{C}})}
{1+\frac{9}{4}\frac{C_2^0}{Q}A^{-1/3}}\right ]+\Delta r_{np}^{coul}+\Delta r_{np}^{surf},
\eea 

where $I=(N-Z)/A$ is the
relative neutron excess in the nucleus, $I_{\scriptsize{C}}=e^2Z/(20C_2^0R), R=r_0A^{1/3},
r_0 = (\frac{3}{4\pi\rho_0})^{1/3}, \Delta r_{np}^{coul} = -\sqrt{3/5}
(e^2Z)/(70C_2^0)$ is the correction caused by electrostatic repulsion and
$\Delta r_{np}^{surf} = \sqrt{3/5}[5(b_n^2-b_p^2)/2R]$, a correction
coming from the difference in surface widths of the neutron and proton
density profiles \cite{Myers1980}.  Expanding Eq.  (\ref{xeq47})to first
order in $\frac{C_2^0}{Q}A^{-1/3}$, a little algebra 
\cite{RocaMaza2013} gives the following relation,
\bea
\label{xeq48}
\alpha_D^{DM} \approx \frac{\pi e^2}{54}\frac{A<r^2>}{C_2^0}[1+\frac{5}{2}
\frac{\Delta r_{np}^{DM}-\Delta r_{np}^{coul}-\Delta r_{np}^{surf}}
{<r^2>^{1/2}(I-I_{\scriptsize{C}})}].
\eea

With an adopted value of $C_2^0=31\pm 2$ MeV, one finds for $^{208}$Pb
that $I_{\scriptsize{C}} \approx 0.028\pm 0.002, \Delta r_{np}^{coul}\approx
-0.042\pm 0.003$ fm. It was further shown for $^{208}$Pb from a large
number of EDFs that $\Delta r_{np}^{\rm surf}\approx 0.09\pm 0.01$
fm\cite{Centelles2010}. Consequently, the small variations in $\Delta
r_{np}^{\rm coul}, \Delta r_{np}^{\rm surf}$ and $I_{\scriptsize{C}}$ can reasonably
be ignored, resulting in, to a good approximation,

\bea
\label{xeq49}
\alpha_D^{DM} \approx \frac{\pi e^2}{54}\frac{A<r^2>}{C_2^0}[1+\frac{5}{2}
\frac{\Delta r_{np}^{DM}}{I<r^2>^{1/2}}].
\eea

\begin{figure}
\centering
\resizebox{0.75\columnwidth}{!}{%
\includegraphics{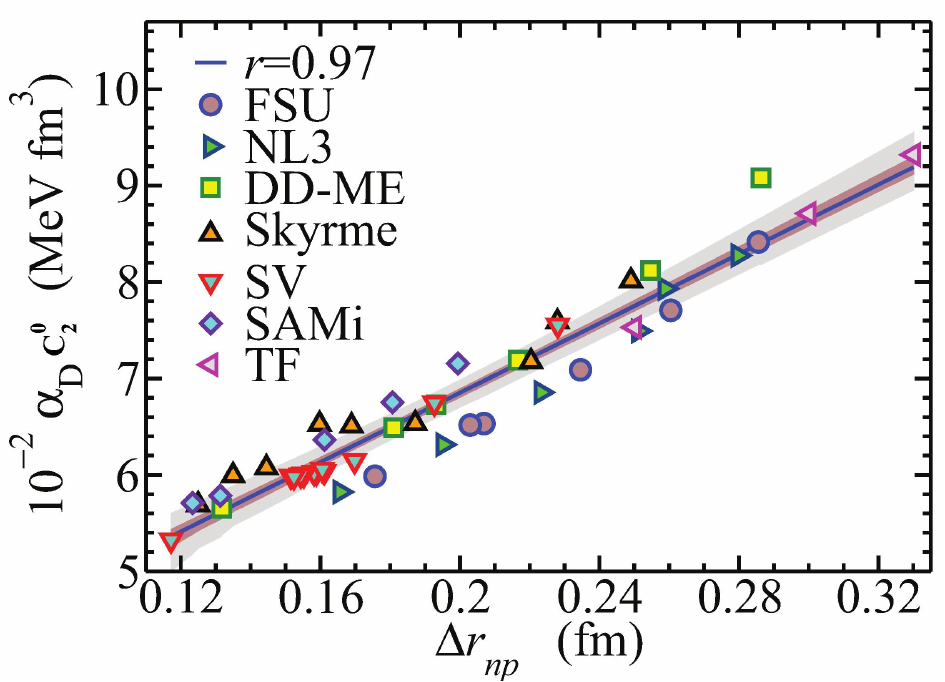} }
\caption{Dipole polarizability times the symmetry energy at
saturation of each model against the neutron skin thickness in 
$^{208}$Pb predicted by the EDFs of Fig. \ref{fig:7}. The
linear fit gives $10^{-2} \alpha_{D} C_2^0=(3.01 \pm 0.32)+(19.22 \pm 0.73)
\Delta r_{\mathrm{np}}$ with  a correlation coefficient $r=0.97,$ and the
two shaded regions represent the $99.9 \%$ and $70 \%$ confidence bands.
This figure is taken from Ref. \cite{RocaMaza2013a}.}
\label{fig:8}       
\end{figure}

Eq. (\ref{xeq49}) suggests a strong correlation between $\alpha_D C_2^0$
with the neutron skin.  This tight correlation (see Fig. \ref{fig:8})
was validated \cite{RocaMaza2013a} in a self-consistent mean-field plus
RPA calculation for both neutron-skin thickness and electric dipole
polarizability using a large set of representative non-relativistic and
relativistic models. If the symmetry energy coefficient $C_2^0$ 
and $\alpha_D$ are known in good precision, a
consequence of this correlation is a pointer to the 
value of the neutron-skin thickness.
 Combining the measured value
of $\alpha_D$ \cite{Tamii2011} for $^{208}$Pb with the adopted value of
$C_2^0$ as mentioned, the neutron skin thickness of $^{208}$Pb is predicted
to be,

\bea
\label{xeq50}
\Delta r_{np}=0.165\pm(0.009)_{exp}\pm(0.013)_{theo}\pm(0.021)_{est} {\rm fm}.
\eea

It is to be noted that in the DM model $\alpha_D C_2^0$ is better correlated to $\Delta
r_{np}$ rather than $\alpha_D$ alone to $\Delta r_{np}$\cite{RocaMaza2013a}; 
this is  in
contravariance to that obtained in Refs. \cite{Tamii2011,Reinhard2010}. 

\subsection{The isovector and isoscalar mass: nucleon isovector mass splitting}

Experimental data on dipole polarizability \cite{Birkhan2017,Hashimoto2015,Tamii2011,Rossi2013}
add new to the wealth of information on atomic nuclei and can be exploited to gain more
confidence  on the nuclear matter  parameters entering 
the EoS. In the following, we show how it
aids in guiding to the nuclear matter  EoS, to 
finding the isovector nucleon mass , the isovector
splitting of nucleon mass and then the other nuclear matter  parameters 
\cite{Malik2018a}. 
The isovector nucleon
mass $m_{v,0}^*$ is the effective mass of a proton in pure neutron matter or vice versa,
and is defined as 
\bea
\label{xeq51}
\frac{m}{m_{v,0}^*}=1+\frac{m}{2\hbar^2}\rho_0 \Theta_v,
\eea

where 
\bea
\label{xeq52}
\Theta_v=\frac{\hbar^2}{2m}(k_+ - k_-).  
\eea
We have already seen  that $k_+$ gives the
isoscalar nucleon mass and $k_-$ (from Eq. (\ref{xeq37})) 
defines the isovector mass splitting. In Skyrme
methodology, one can check from Eq. (\ref{xeq34}) 
that the isovector parameter $\Theta_v$ is,
\bea
\label{xeq53}
\Theta_v=t_1(1+x_1/2)+t_2(1+x_2/2).
\eea 

The energy weighted sum rule $m_1$ (see Eq. (\ref{xeq40}) 
for the isovector giant dipole
resonance (IVGDR) of a nucleus can be written as \cite{Bohr1975}, 
\bea
\label{xeq54}
m_1=\frac{9}{4\pi}\frac{\hbar^2}{2m}
\frac{NZ}{A}(1+\kappa_A),
\eea
where $\kappa_A$ is the polarizability enhancement factor 
for the nucleus. It has a
relation to $\Theta_v$ as \cite{Chabanat1997}, 
\bea
\label{xeq55}
\kappa_A=\frac{2m}{\hbar^2}\frac{A}{4NZ}\Theta_v\times I_A,
\eea
where the integral $I_A=\int\rho_n(r)\rho_p(r)d^3r$, $\rho_n(r)$ and $\rho_p(r)$ being the
neutron and proton density distributions 
in the nucleus. In principle, $m_1$ can be determined
from the experimental strength function $R(\omega)$, $\kappa_A$ is then determined. If
$I_A$ is known from some other source, then $\Theta_v$, and hence the isovector mass can
be calculated. If the isoscalar effective mass is further known, one gets $k_-$ and thence
the isovector mass splitting $\Delta m_0^*$. Any knowledge of the dipole polarizability
$\alpha_D$ is redundant in the extraction of $\Theta_v$, it appears. 

The fact that the high energy component of the strength function is
plagued  with 'quasi-deuteron effect' renders the determination of $m_1$
not very reliable; this forces us to look into dipole polarizability as an
extra experimental input. From Eqs. (\ref{xeq38}) and (\ref{xeq39}), $m_1$ is written as, 
\bea
\label{xeq56}
m_1=\frac{9}{8\pi e^2}E_x^2 \alpha_D.
\eea
  To find the values of $\Theta_v$, values of $m_1$ are constructed
from reasonable inputs on the constrained energy $E_x$ and the integrals
$I_A$, which are found from correlation systematics.

\begin{figure}
\centering
\resizebox{0.75\columnwidth}{!}{%
\includegraphics{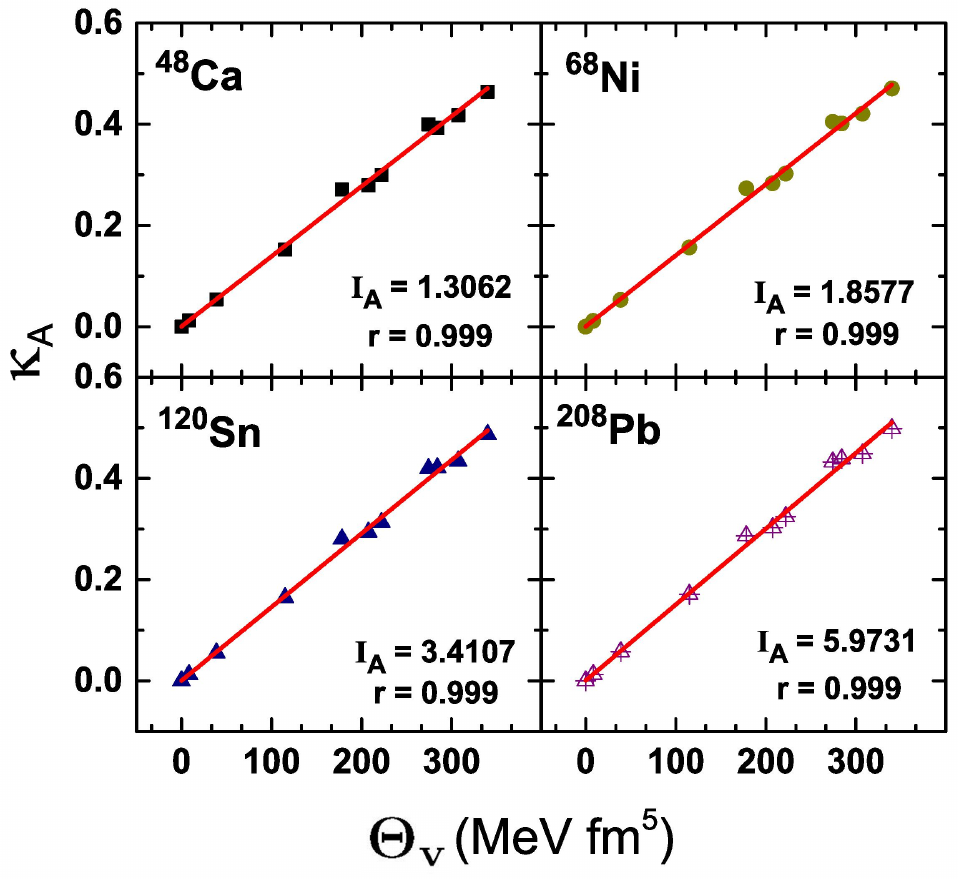} }
\caption{The correlation of the isovector parameter $\Theta_{v}$
obtained from the Skyrme EDFs \cite{Brown2013} with the calculated
dipole enhancement factor $\kappa_{A}$ for the nuclei $^{48}
\mathrm{Ca},^{68} \mathrm{Ni}, ^{120} \mathrm{Sn}$, and $^{208}
\mathrm{Pb}$. The corresponding values of the integrals $I_{A}$ (in units
of $\mathrm{fm}^{-3}$ ) and the correlation coefficients are shown in
each panel. The figure is taken from Ref. \cite{Malik2018a}.}

\label{fig:9}       
\end{figure}

We first find the isovector  integrals $I_A$. From the neutron and proton densities $\rho_n(r)$ and
$\rho_p(r)$ calculated in the Hartree-Fock (HF) 
approximation for the four nuclei, viz, $^{48}$Ca,
$^{68}$Ni, $^{120}$Sn and $^{208}$Pb (for which the data for $\alpha_D$ are available) with selected
Skyrme EDFs (called 'best-fit' Skyrme EDFs \cite{Brown2013}), it is found that the integrals $I_A$ for a
particular nucleus are nearly independent of EDFs \cite{Malik2018a}. 
The values of $m_1$ calculated from the HF +RPA and
hence $\kappa_A$ are different for different EDFs with differing values of $\Theta_v$, but there is an
extremely strong correlation (with correlation coefficient $r$ practically unity) between $\Theta_v$ and
$\kappa_A$ as displayed in Fig. \ref{fig:9}. 
The slopes of the correlation lines are taken as measures
for $I_A$ for each nuclei; they are shown in respective panels in the figure. 

As noted already, experimental values of $m_1$ are 
somewhat uncertain due to 'quasi-deuteron'
effect. With reasonable choice of the constraint energy $E_x$, 
we try to gauze  $m_1$ in
good bounds from existing data on $\alpha_D$ for the four nuclei as mentioned. Two choices of $E_x$ are
made. For the lower values of $E_x$, the known peak energy $E_p$ of the experimental IVGDR strength
function is chosen. For the higher value, 
$E_x = 1.05E_p$ is taken. This choice is not arbitrary, in RPA
calculations with the 'best-fit' Skyrme EDFs\cite{Brown2013}, it has always been found that $E_x$ is
higher than $E_p$ by $\sim (4 - 6)\%$ for the nuclei studied. These two choices of $E_x$ provide the
lower and upper bounds of $m_1$. 

\begin{figure}
\centering
\resizebox{0.75\columnwidth}{!}{%
\includegraphics{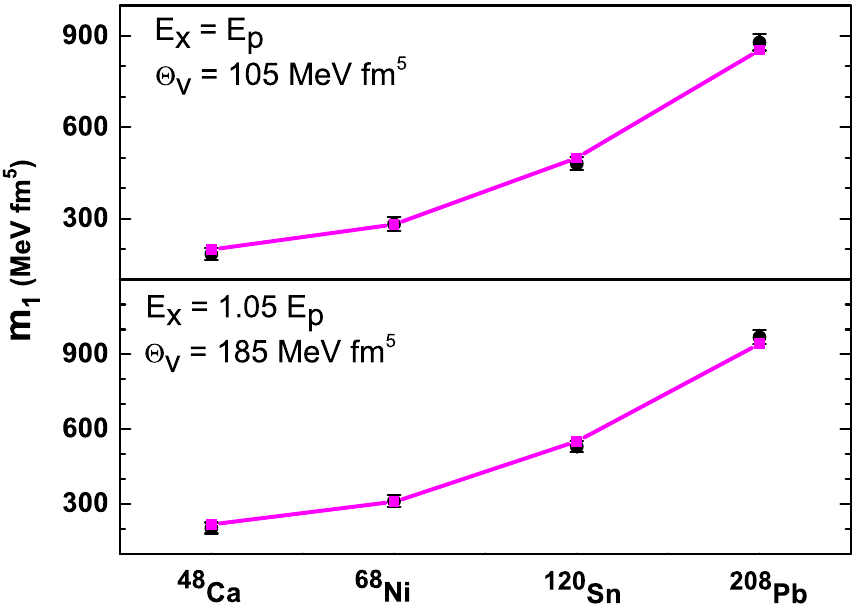} }
\caption{The calibrated values of EWSR $\left(m_{1}\right)$ displayed for
the four nuclei (black squares). The upper panel shows the values with
$E_{x}=E_{p},$ the lower panel displays the same with $E_{x}=1.05 E_{p}$
(see text for details). The solid lines are drawn to show the fit with
$\Theta_{v}=105 ~\mathrm{MeV} \mathrm{fm}^{5}$ and $185 ~\mathrm{MeV}
\mathrm{fm}^{5},$ respectively. The figure is taken from Ref.
\cite{Malik2018a}.}

\label{fig:10}       
\end{figure}
The value of $\Theta_v$ is then obtained as follows: $E_x$ equated with
$E_p$, $m_1$ calculated from Eq. (\ref{xeq56}) for the four nuclei
with the experimental values of $\alpha_D$, $\kappa_A$ obtained from
Eq. (\ref{xeq54}). With known values of $I_A$, the so-obtained $\kappa_A$
are then subjected to a $\chi^2$ minimization by varying $\Theta_v$
(Eq. (\ref{xeq55})).  The optimized value of $\Theta_v$ is found to be
$\Theta_v = 105.0$ MeV fm$^{5}$. The calculation is repeated with $E_x =
1.05 E_p$. The optimized value of $\Theta_v$ is now 185.0 MeV fm$^{5}$.
The fitted two sets of results are shown in Fig. \ref{fig:10}.
The fits are seen to be very good in both cases. An average value of
$\Theta_v\approx 145.0\pm 40.0$ MeV fm$^5$ can be inferred from the
calculation. Since, $\Theta_v$ is the difference between $k_+$ and $k_-$
and is a constant, if $k_+$ increases $k_-$ should also increase or
vice versa. The dipole polarizability then helps to find the isovector
parameter $\Theta_v$ or the isovector mass $m_{v,0}^*$, but does not give
directions to separately find $k_+$ (the isoscalar mass $m_0^*$) or $k_-$
(the isovector mass splitting $\Delta m_0^*$).

Inference on the value of the effective isoscalar mass $m_0^*$
has been drawn from many corners; they seem to lie in a rather
broad range. Skyrme EDFs yield $m_0^*/m$  in the range $0.6 - 1.0$
\cite{Brown1980,Chen2009,Dutra2012,Davesne2018}. Many body calculations,
irrespective of their level of sophistication give $m_0^*/m\sim 0.8\pm
0.1$\cite{Friedman1981,Wiringa1988,Zuo1999}. Analysis of isoscalar giant
quadrupole resonance (ISGQR) \cite{Stone2007,RocaMaza2013,Zhang2016}
points to a similar value $(\sim 0.85\pm 0.1)$, but the analysis is
model dependent.  Optical model analyses of nucleon-nucleus scattering,
on the other hand, yield a value of the effective mass somewhat less,
$m_0^*/m\sim 0.65\pm 0.06$ \cite{Li2015}, relativistic models compatible
with saturation properties of nuclear matter together with the constraints
on low density neutron matter from chiral effective field theory {\it
ab-initio} approaches \cite{Drischler2016} and some recent astrophysical
constraints also yield a value of effective mass in a similar range,
$0.55\le m_0^*/m\le 0.75$ \cite{Hornick2018}.

 \subsection{ Fitting nuclear macro data: approaching an EoS}

\begin{figure}
\centering
\begin{tabular}{c}
  \resizebox{0.62\columnwidth}{!}{%
\includegraphics{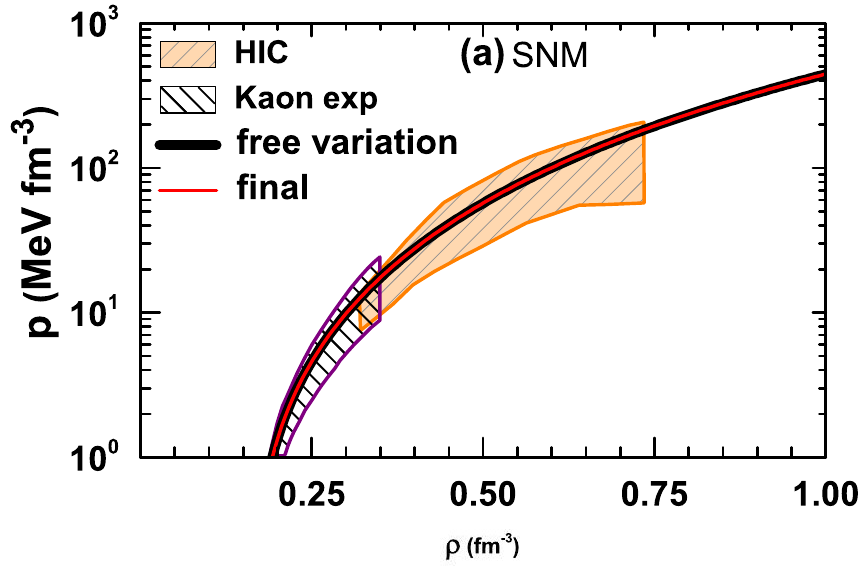}}\\ 
\resizebox{0.62\columnwidth}{!}{\includegraphics{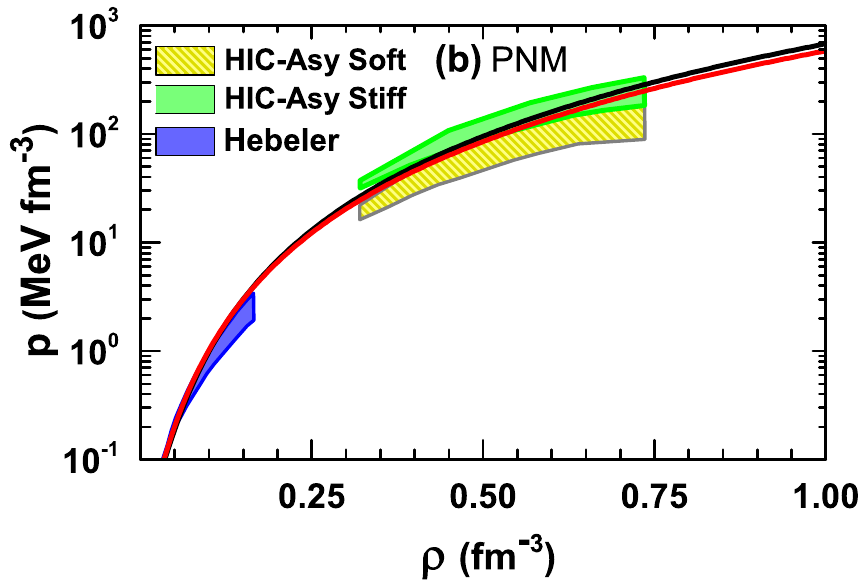}}\\
 \resizebox{0.62\columnwidth}{!}{\includegraphics{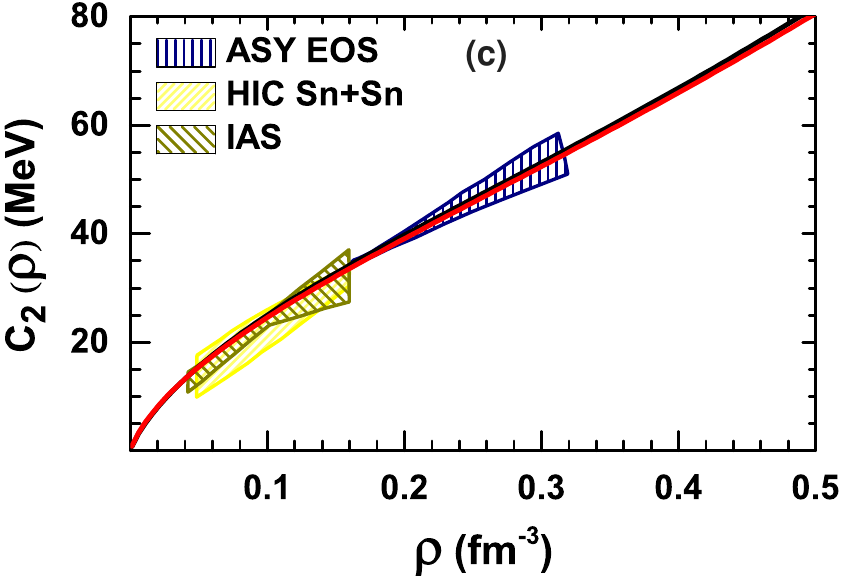}}
\end{tabular}
\caption{The pressure $P(\rho)$ for SNM (top), PNM (middle) 
and symmetry energy coefficient
$C_2(\rho)$ (bottom) as a function of baryon density $\rho$ for the 
best-fit parameters obtained from free variation of all the parameters 
(black lines) and for the final values of the parameters (see Table \ref{tab2}) 
shown by the red line. The figure is taken from Ref. \cite{Malik2018a}.
}
\label{fig:11}       
\end{figure}

All the seven model  parameters appearing in Eq.(\ref{xeq34}) that enter
in the nuclear matter  EoS can be determined, in principle, in a single go from a
$\chi^2$-fitting of the nuclear 'macro data'. By macro data, we mean data
on nuclear matter  pressure, energy and symmetry energy at different densities
accumulated from different experiments involving nuclear collisions and
subtle theoretical arguments. They are listed in Table \ref{tab1}.     The fitting
protocol in addition, includes values of empirical nuclear matter  parameters
pertaining to SNM, namely its energy per nucleon $e_0$, the saturation
density $\rho_0$ and the incompressibility $K_0$. A free variation of all
the parameters yields a very shallow minimum in $\chi^2$ corresponding
to $m_0^*/m \approx 1.31$ \cite{Malik2018a}.  To get an insight into
this flatness problem, we constrain $\tilde\alpha$ to a fixed value
and then optimize $\chi^2$ varying the remaining six parameters. Each
choice of $\tilde \alpha$ leads to a different set of EDF parameters
and hence $m_0^*/m$. Each  parameter set is found to be equally good
in fitting the macro data (see Fig. \ref{fig:11}).  An unique value of
$m_0^*/m$ can not thus be arrived at from this fitting. We, however,
find an unique relationship between $m_0^*/m $ and $\Delta m_0^*/\delta$
in this fitting procedure, it is found that  $\Delta m_0^*/\delta$
decreases with increasing $m_0^*/m$ in an almost parabolic way that can
be well approximated as

\bea
\label{xeq57}
\Delta m_0^*/\delta=\beta_1+\beta_2(m_0^*/m)^2,
\eea

with $\beta_1 = 0.733\pm 0.024$ and $\beta_2 = -2.029\pm 0.032$. 
This equation can be restated as

\bea
\label{xeq58}
(k_-+\beta_1k_+)\rho_0 \approx -(\beta_1+\beta_2).
\eea

\begin{figure}
\centering
\resizebox{0.75\columnwidth}{!}{%
\includegraphics{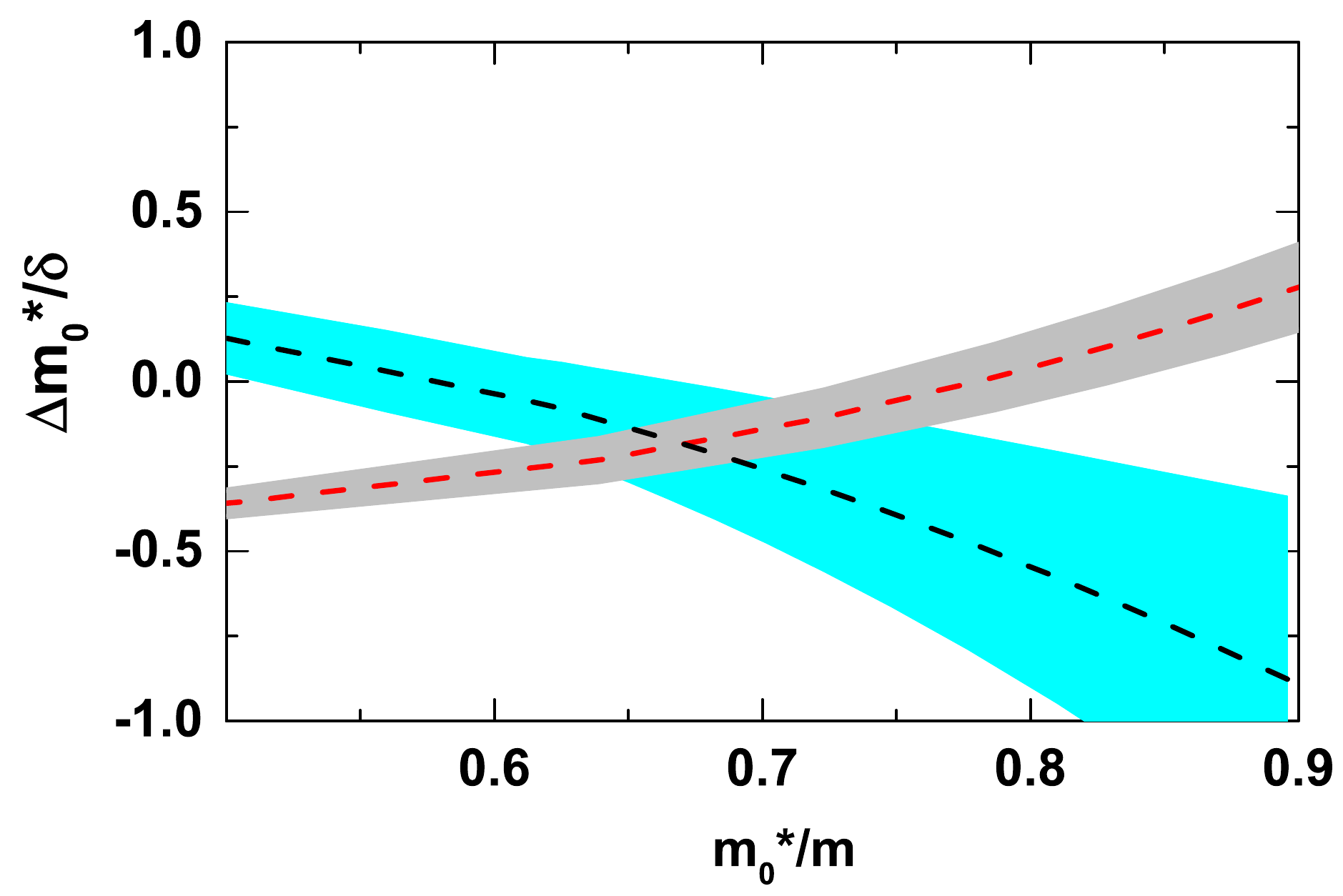} }
\caption{The isovector effective mass-splitting as a 
function of nucleon effective mass. The black dashed line 
refers to the best fit obtained from macrodata for 
different values of $\tilde \alpha$; the red dashed line 
corresponds to the one obtained by satisfying 
Eq. (\ref{xeq57}) with $\Theta_v = 145$ MeV fm$^5$. The
cyan and grey shades refer to the respective uncertainties.
The figure is taken from Ref. \cite{Malik2018a}.}
\label{fig:12}       
\end{figure}

Since, $\beta_1$ and the r.h.s of this equation  is seen to be positive,
one finds that as $k_-$ increases, $k_+$ decreases and vice versa. Coupled
with the opposing trend obtained from dipole polarizability  where
$k_-$ increases as $k_+$ increases, one sees (see Fig. \ref{fig:12})
that unique values of $m_0^*/m$ and then $\Delta m_0^*/\delta$ can be
obtained. The value of effective mass is $0.61\le m_0^*/m \le 0.75$
and for isovector splitting, $-0.3\le \Delta m_0^*/\delta \le -0.1$. The
final model parameters of the EoS \cite{Malik2018a} are listed in  Table  \ref{tab2}  
that contains the correlated and uncorrelated errors obtained within the
covariance method. The isovector mass comes out to be $m_{v,0}^*/m =
0.78^{+0.05}_{-0.04}$. The other nuclear matter  parameters can be calculated
with the EoS, they are listed in Table \ref{tab3}

\section{ Message from the heavens}
\subsection{Neutron star properties: relations to symmetry elements}

Neutron stars are incredibly dense objects, made of baryonic matter,
mostly of neutrons. Matter at supra-nuclear densities, as encountered
in the core of the neutron star can not be accessed in terrestrial
experiments, astrophysical observations involving the neutron stars are
thus essential in understanding dense matter EoS. {\bl The degenerate baryon
pressure balances the gravitational pull preventing stars  as  heavy as of 
mass $\approx 2M_\odot$ from turning into black-holes.}  The so-far observed
maximum mass of the neutron star with $M_{\rm NS} \approx 2M_\odot$
\cite{Antoniadis2013,Demorest2010,Rezzolla2018,Cromartie2019} thus serves
as a stringent constraint on the nuclear matter  EoS. This  observation sets, till
now, the absolute limit on the softness/stiffness of the EoS and helps
in focussing on those EDFs that satisfy this observational criterion.

\begin{figure}
\centering
\resizebox{0.750\columnwidth}{!}{%
\includegraphics{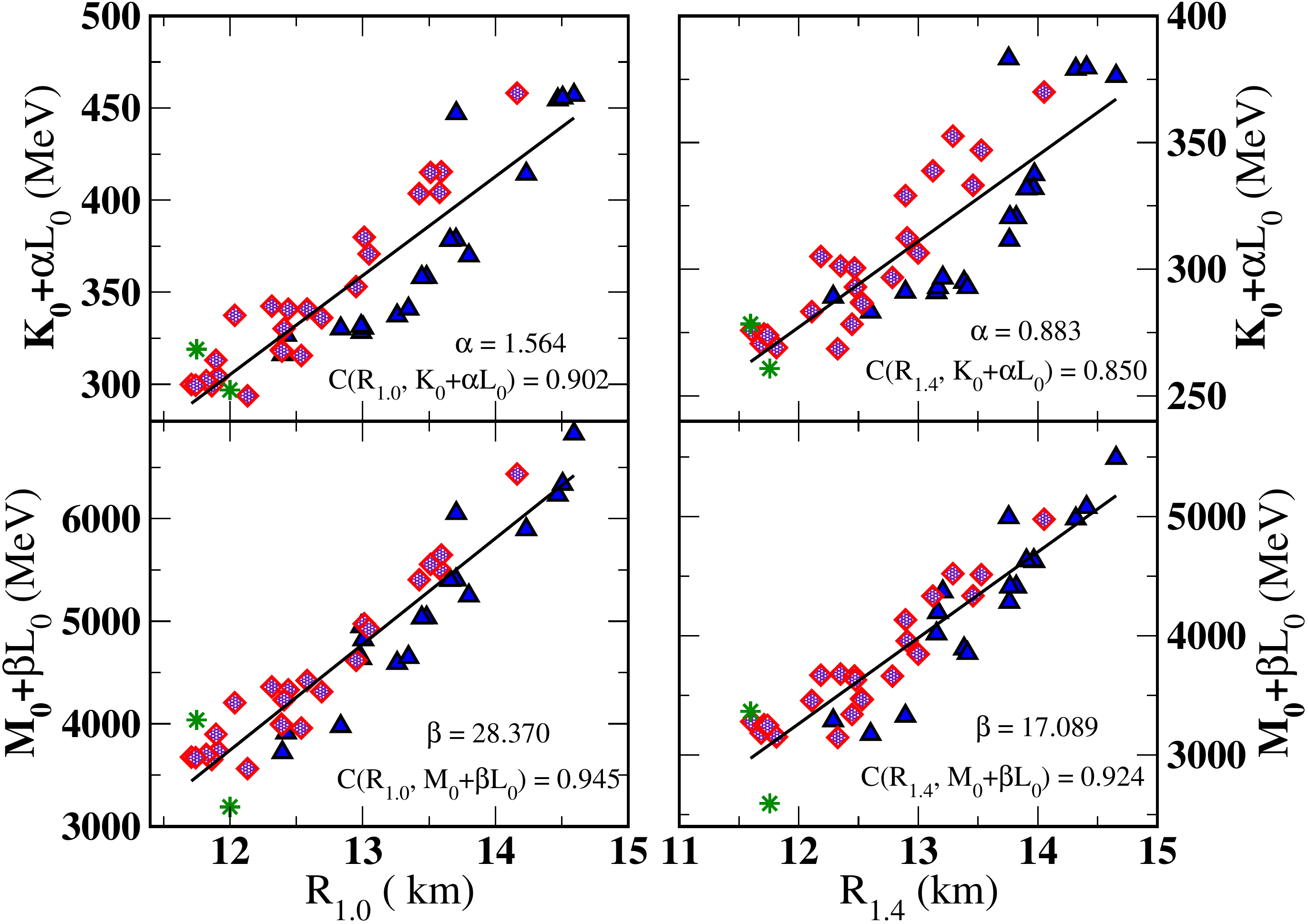} }
\caption{Neutron star radii $R_{1.0}$ (left) and $R_{1.4}$ (right) 
versus the linear correlations $K_{0}+\alpha L_{0}$ ( top ) 
and  $M_{0}+\beta L_{0}$ (bottom), using a set of 
RMF (blue triangles), Skyrme (red diamonds), and 
BHF+APR (green stars) calculations. The figure is taken from 
Ref. \cite{Alam2016}.}
\label{fig:13}    
\end{figure}

Since, the EoS is determined, in principle, by Eqs. (\ref{xeq3}) and
(\ref{xeq4}), the nuclear matter  parameters entering the EoS should show
some correlation  with some properties of neutron star,
such as the crust-core transition density and pressure, radii,
maximum mass etc. As examples, the crust-core transition density
is seen to be strongly correlated to the symmetry slope $L_0$ or
the neutron skin of a heavy nucleus \cite{Vidana2009,Ducoin2010},
the transition pressure is found to be strongly correlated with a
linear combination of $L_0$  and the symmetry curvature $K_{\rm sym}^0$
 at a sub-nuclear density ($\rho\sim 0.1$ fm$^{-3}$)
\cite{Ducoin2011,Newton2013,Fattoyev2014}. Such correlations point
to some general interrelationships between the nuclear matter  parameters
and the properties of high density nuclear matter. They may only be
useful in having a close understanding of a still largely unknown
nuclear matter parameters  when its interrelating partners are well known. For
example, the simultaneous determination of mass and radius of a low
mass neutron star has been shown to constrain better the product of
nuclear matter  incompressibility $K_0$ and the symmetry slope $L_0$\cite
{Sotani2015} but till now, it appears, the NS radius or $L_0$ are not
that well constrained; to be more specific, if the mass and radius of
the low mass NS and $K_0$ are to be taken to the constrained values,
the value of $L_0$ seems in present knowledge uncharacteristically
large. It may not be hard to comprehend that low mass neutron stars ($\sim
0.6M_\odot$) may not involve too high core density for degenerate baryon
pressure to support the gravitational pull; with that understanding,
the radii of low mass neutron stars were correlated with the neutron
skin of $^{208}$Pb (a manifestation of neutron pressure at $\sim \rho_0$
\cite{Fattoyev2010}). {\bl However neutron star with such low mass are
not yet discovered, the lowest NS mass observed so far is $1.17\pm 0.004 
M_\odot$ \cite{Martinez2015}.} It was found that the correlation is extremely strong for
low mass neutron stars, getting weaker as the mass of the neutron star
increases.  In the same vein, Alam {\it et al} \cite{Alam2016} investigate
the correlation of neutron star radii with the key nuclear matter
parameters governing the nuclear matter  EoS; they find a correlation of
the NS radii with $K_0, M_0,$ and $L_0$ but not strong enough to give
a good understanding of the EoS. A linear combination of these nuclear
matter  parameters like $K_0+\alpha L_0$ and $M_0+\beta L_0$ with the NS
radii give a better correlation, the correlation becoming stronger with
lower masses of the neutron star (see Fig. \ref{fig:13}).  Though, the
values of $L_0$ and $M_0$ are not very certain, their plausible values
as deduced from finite nuclear data constrain the radius $R_{1.4}$
of a canonical star of mass $1.4M_\odot$ in the range 11.09 - 12.86 km.

\subsection{Tidal deformability and relations to symmetry elements}

In August 2017, the advanced LIGO and advanced VIRGO gravitational 
wave observatories detected
gravitational waves (The GW170817 event) 
from merger of two neutron stars \cite{Abbott2017}. 
During the last stages of the
inspiral motion of the coalescing neutron stars the strong gravity of each induces a strong tidal
deformation in the companion star. The gravitational wave phase evolution caused by the deformation
\cite{Flanagan2008} is decoded allowing for the determination of a dimensionless tidal deformability
parameter $\Lambda$ \cite{Hinderer2008,Hinderer2010,Damour2012}. It measures the response of the gravitational
pull on the surface of the neutron star and 
thus becomes the correlator of  the pressure gradients inside
the NS serving as an effective probe of the high density nuclear matter  EoS \cite{Hawking1987,Read2009}. A
relatively large value of $\Lambda$, for example, 
points to a relatively large neutron star
radius\cite{Annala2018,De2018,Malik2018}; that speaks of 
a stiff EoS and hence a comparatively large
value of the neutron skin of a heavy nucleus \cite{Fattoyev2018}.

The tidal deformability parameter $\lambda$ is defined as \cite{Flanagan2008,Hinderer2008,Damour2012}
\bea
\label{xeq59}
Q_{ij}=-\lambda \mathcal{E}_{ij},
\eea

where $Q_{ij}$ is the induced quadrupole moment of a star in a binary due to the static external tidal
field $\mathcal{E}_{ij}$ of the companion star. 
The parameter $\lambda$ can be expressed in terms of the
dimensionless quadrupole tidal Love number $k_2$ as (we take the 
geometrized unit $G=c=1$),
\bea
\label{xeq60}
\lambda=\frac{2}{3}k_2R^5,
\eea

where $R$ is the radius of the NS. The value of $k_2$ depends on
the stellar structure, it can be obtained \cite{Hinderer2008,Malik2018}
in conjunction with solving for the Tolmann-Volkoff equations
\cite{Weinberg1972}. Typically, the value of $k_2$ lies in the range
$\approx 0.05 - 0.15$ \cite{Hinderer2008,Postnikov2010} for neutron
stars.   The dimensionless tidal deformability  
is then defined as $\Lambda=\frac{2}{3}k_2 C^{-5}$
where $C (\equiv M/R)$ is the compactness of the star of mass $M$. The Love
number  $k_2$ has a veiled NS radius dependence: Ref. \cite{De2018} finds
$\Lambda\sim R^6$, Ref. \cite{Tsang2019} finds it as $\sim R^{6.26}$
while in Ref. \cite{Malik2018}, for a NS of canonical mass $1.4M_\odot$,
it is $\sim R^{6.13}$. There is thus expected a strong correlation between
$\Lambda$ and $R$. The tidal deformabilities of the neutron stars present
in the binary system can be combined to yield an weighted average as 
\bea
\label{xeq61}
\tilde \Lambda =\frac{16}{13} \frac{(12q+1)\Lambda_1+(12+q)q^4\Lambda_2}
{(1+q)^5},
\eea

where $\Lambda_{1,2}$ are the tidal deformabilities of the NSs of mass
$M_1$ and $M_2$ and $q=M_2/M_1\le 1$ is the binary's mass ratio. Early
analysis of the GW170817 event \cite{Abbott2017} puts an upper
limit for $\tilde\Lambda$ at $\approx 800$ for the component neutron
stars with masses in the range $\approx 1.17 M_\odot - 1.6M_\odot$
involved in the merger event. Revised values of $\tilde\Lambda$ seem
to be substantially lower \cite{De2018,Abbott2018,Abbott2019}. With a
few plausible assumptions for a canonical neutron star, a restrictive
constraint has been set for $\Lambda_{1.4}$ to $\sim 190^{+390}_{-120}$
\cite{Abbott2018}. From the spectral parameterization of the pressure
$P(\rho)$ for the $\beta$-equilibrated matter to fit the observational
template, the pressure inside the NS at supra-normal densities is also
predicted. The goal now remains to be seen: to find the most realistic
EoS that connects all the constraints involving microscopic nuclei with
those obtained from astrophysical scenario, namely, the maximum neutron
star mass and the tidal deformability.

\begin{figure}
\centering
\resizebox{0.95\columnwidth}{!}{%
\includegraphics{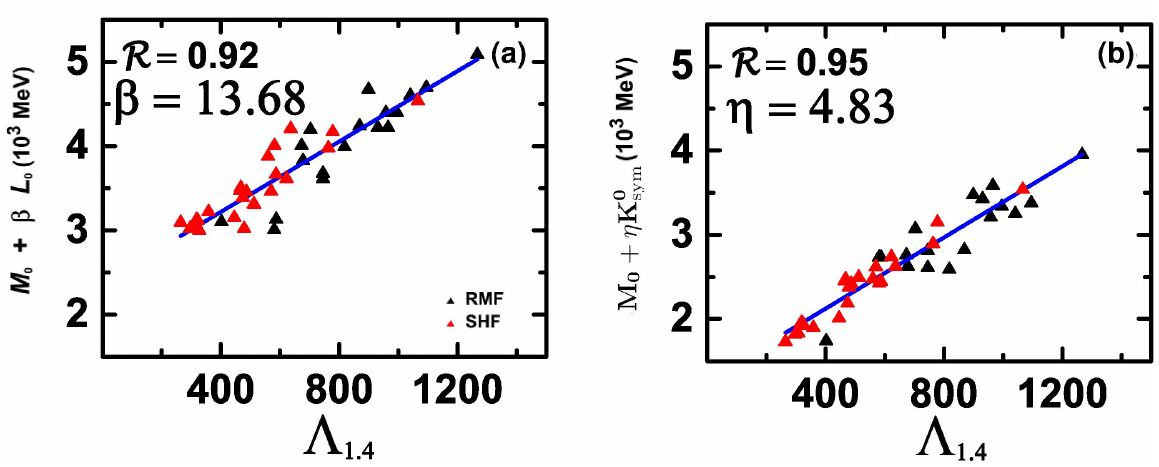} }
\caption{(a) The $M_0+\beta L_0$ and (b) $M_0+\eta K_{\rm sym}^0$ 
                             versus the dimensionless tidal deformability 
                             $\Lambda_{1.4}$ for a  
                             $1.4 ~ {\rm M}_{\odot}$ NS, 
using a set of RMF and SHF models. The figure is taken from
Ref. \cite{Malik2018}.}
\label{fig:14}       
\end{figure}

\begin{figure}
\centering
\resizebox{0.95\columnwidth}{!}{%
\includegraphics{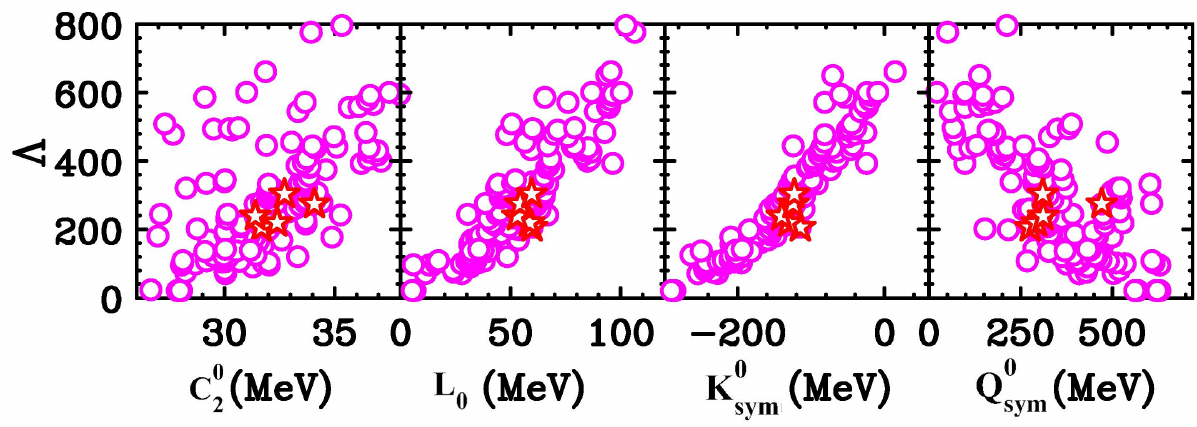} }
\caption{The four panels show the correlation between the tidal
deformability of  neutron star$\Lambda$ for 1.4 solar-mass neutron stars
and Taylor expansion coefficients (from left to right)  $C_2^0 , L_0,
K_{\rm {\rm sym}}^0 {\rm and~} Q_{\rm {\rm sym}}^0$ defined in Eq. \ref{xeq4}
obtained for the Skyrme functionals used in Ref. (\cite{Tsang2019}). The
figure is taken from Ref. \cite{Tsang2019}} \label{fig:15}
\end{figure}

Initial attempts in this direction have been made on understanding the
sensitivity of the tidal deformability to the nuclear matter  parameters related
to nuclear matter at saturation density. In Ref.  \cite{Malik2018}, the
correlation of $\Lambda, k_2$ and $R$ with the nuclear matter  parameters $K_0,
Q_0, M_0, C_2^0, L_0, K_{\rm sym}^0$ and with several linear combinations
of two parameters, in particular, $K_0+\alpha L_0, M_0 +\beta L_0$
and $M_0+\eta K_{\rm sym}^0$ are studied with a set of 18 relativistic
and 24 non-relativistic nuclear models that are known to yield a good
description of the properties of finite nuclei and neutron stars.  
The correlation systematics is determined for 
NS masses in the  range $1.2 M_\odot - 1.6 M_\odot$, since for analysis
of low spin prior as assumed in Ref. \cite{Abbott2017}, these masses
are close to the GW170817 event. Calculations in Ref. \cite{Malik2018}
show that the individual nuclear matter  parameters are weakly or moderately
correlated with $\Lambda, k_2$ and $R$, but $\Lambda$ and $R$ have tight
correlation with $M_0+\beta L_0$ and $M_0 +\eta K_{\rm sym}^0$ over a
wide range of  NS masses considered; the correlation coefficient $r$ is
$\sim 0.9$. The Love number $k_2$ is, however, only strongly correlated
with $M_0 + \eta K_{\rm sym}^0 (r\sim 0.92)$.  The values of $\alpha,
\beta$ and $\eta$ are obtained from the demand of optimum correlations
for each NS mass; they are found to decrease monotonically with increase
in NS mass. This indicates that the density dependence of the symmetry
energy is less important in determining the tidal deformability and
the radius at higher NS masses.  Representative examples of the fit of
$M_0 +\beta L_0$ and $M_0 + \eta K_{\rm sym}^0$ with $\Lambda_{1.4}$
are shown in Fig. \ref{fig:14}.
The figure shows that once $\Lambda_{1.4}$ and $L_0$ are known within
tight limits, $M_0$ can be constrained and then $K_{\rm sym}^0$. Empirical
values of $M_0$ and $K_{\rm sym}^0$ derived for different limits of
$\Lambda_{1.4}$ and $L_0$ are shown in Table \ref{tab4}. On the
other end, the strong correlation of $\Lambda_{1.4}$ with $R_{1.4}$
mentioned earlier puts a strong constraint on the radius of a
canonical neutron star that can be compared to that obtained from
the simultaneous determination of the radius and mass of a NS
with the NICER (Neutron star Interior Composition Explorer)
mission \cite{Riley2019,Raaijmakers2019,Miller2019}. To further the
understanding of the relationship of the tidal deformability to the
isovector nuclear matter parameters, Tsang et. al \cite{Tsang2019}
studied, with a total of 240 Skyrme interactions the correlation of
$\Lambda $ with $C_2^0,~L_0,~K_{\rm sym}^0$ and $Q_{\rm sym}^0$; they found little
correlation except between $\Lambda -L_0$ and $\Lambda -K_{\rm sym}^0$, the
later one  being the strongest  as displayed in  Fig.\ref{fig:15}.
In Ref.\cite{Zhang2018}, a stronger correlation between $\Lambda $ and
$L_0$ than between $\Lambda $ and $K_{\rm sym}^0$ is reported; this has to be
critically examined further as one expects the opposite since $K_{\rm sym}^0$
impacts higher densities more.

To avoid model dependence on the results of correlations between
nuclear matter parameters and the NS properties and to look
for further correlations, an approach taken in several works
\cite{De2015,Margueron2018a,Margueron2019} termed 'meta-modelling' has
been applied recently \cite{Ferreira2020} to construct the nuclear matter
EoS. Based on the Taylor expansion of the energy functional around
the saturation density (as shown in Eqs.\ref{xeq3} and \ref{xeq4}), with
expansion coefficients identified as different nuclear  matter parameters, the EoS
in principle, is model-independent provided the nuclear matter parameters are
experimentally known.  Exploiting the idea, in Ref. \cite{Ferreira2020},
millions of EOSs are constructed with the eight nuclear matter  parameters
($K_0$, $Q_0$, $Z_0$, $C_2^0$, $L_0$, $K_{\rm sym}^0$, $Q_{\rm sym}^0$,
$Z_{\rm sym}^0$) each one having a Gaussian distribution around the
supposedly known central values. The  multivariate Gaussian is taken
to have zero covariance. The saturation energy and density are fixed at
$e_0 = -15.8$ MeV and $\rho_0$ = 0.155 fm$^{-3}$. Out of the million EoSs
so generated only $\sim $ 2000 are selected to be valid ones filtered
from imposition of the following constraints: (i) the EoS must be
thermodynamically stable (ii) it should be causal, (iii) should support
the observational constraint of the maximum mass at least as high as 1.97
$M_\odot$, (iv) the tidal deformability should be $70 < \Lambda_{1.4} <
580$ \cite{Abbott2017} and (v) the symmetry  energy $C_2(\rho)$ should be
positive. All the EoSs are obtained for the  matter composed of neutrons,
protons, electrons and muons in  $\beta$-equilibrium. The low density part
($\rho < 0.1$ fm$^{-3}$) of the EoSs are matched with the SLy4 EoS so
that $P_{SLy4}(\mu)=P_{EoS}(\mu)$ where $\mu$ is the chemical potential.

\begin{figure}
\centering
\begin{tabular}{c}
\includegraphics{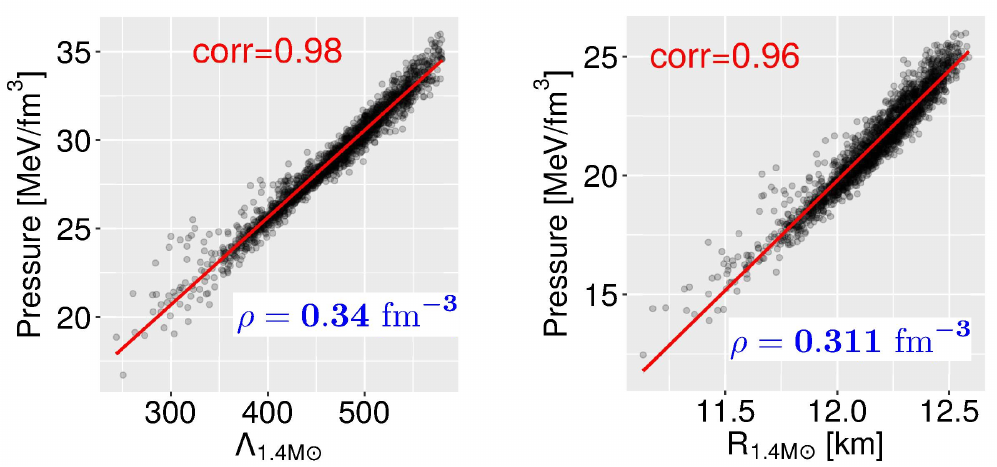}
\end{tabular}
\caption{$\Lambda_{1.4 M_\odot}$ (left) and $R_{1.4 M_\odot}$ (right) as a 
 		  function of the pressure
           for all meta models 
used in Ref. \cite{Ferreira2020} at the 
          densities corresponding to the maximum correlations. These 
          densities and correlation coefficients 
are indicated in each of the panels. The figure is taken from 
\cite{Ferreira2020}.}
\label{fig:16}       
\end{figure}

The filtered EoSs are employed to calculate the mass ($M$) radius
($R$) and the tidal deformability ($\Lambda$) of neutron stars
and then to study the possible existing correlations between the NS
observables and the thermodynamic properties of dense stellar matter in
$\beta$-equilibrium. Strong correlations between them are found to build
up at different densities depending on the NS masses, for example,
correlation between $P(\rho)$  for $\beta$-equilibrated matter  and
$R_M$ (radius of NS of mass M) is found to be almost unity at $\rho\sim
0.2$ fm$^{-3}$ for NS mass of 1.0 $M_\odot$; this peak value shifts to
$\sim \rho = 0.35$ fm$^{-3}$ when the NS mass is 1.6$M_\odot$. A similar
correlation is observed between $P(\rho)$ and $\Lambda_M$ in the same
range of densities. Strong correlations are also identified between the
energy density $\mathcal{E}(\rho)$ with all the NS observables, but   at
larger densities, $\rho\simeq 0.32 - 0.5$ fm$^{-3}$, the smaller NS masses
corresponding to smaller densities. The strong correlation $r\sim 1.0$
can be used as a tool to constrain thermodynamic quantities at different
densities from the NS observables. For example, correlations  between  $P$
and $\Lambda$ and between $P$ and $R$ for NS mass 1.4$M_\odot$
are shown in the left and right panels of Fig. \ref{fig:16}.
\begin{figure}
\centering
\resizebox{0.75\columnwidth}{!}{%
\includegraphics{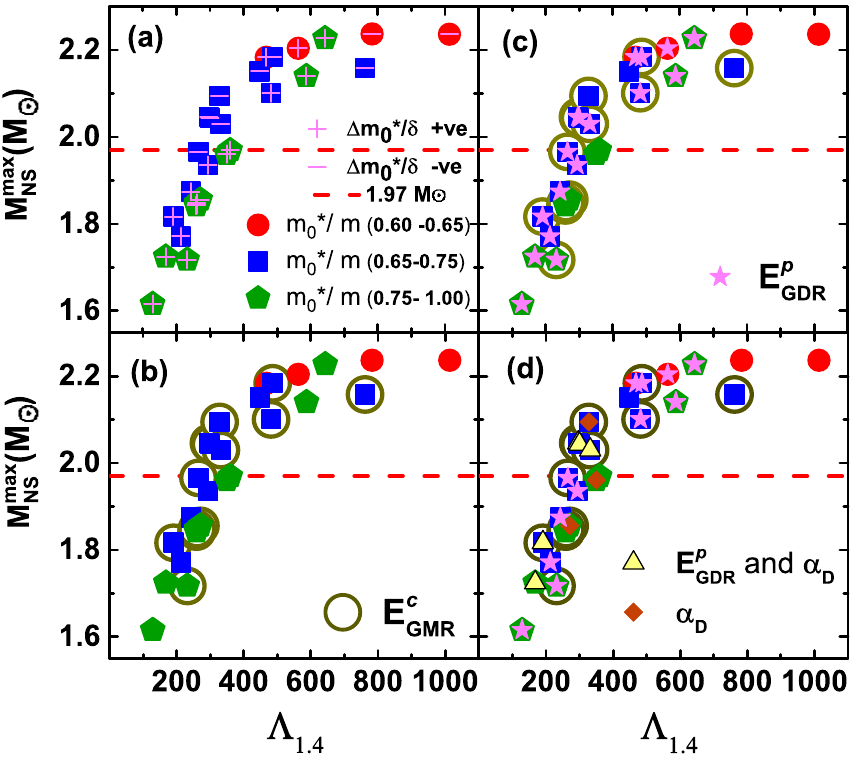} }
\caption{The maximum neutron star mass $M_{\rm
NS}^{\rm max}$ versus the tidal deformability parameter $\Lambda_{1.4}$
obtained from the 28 selected EDFs.  The red dashed lines refer to
$1.97M_\odot$, the observed lower bound for $M_{\rm NS}^{\rm max}$
. For more details , see  Ref. \cite{Malik2019}. The figure is borrowed 
from Ref. \cite{Malik2019}.}
\label{fig:17}       
\end{figure}
In both cases, the correlation is found to be very strong at $\rho\sim
0.32$ fm$^{-3}$ . If a simultaneous precise measurement of the mass
and radius of NS is possible, say with the NICER mission, then from the
radius, a constraint on the pressure of the $\beta$ stable stellar matter
can be obtained at $\rho\sim 0.32$ fm$^{-3}$. This in turn gives an idea
about $\Lambda_{1.4}$. In passing we note that in Ref. \cite{Ferreira2020},
correlations between the energy density $\mathcal{E}$ and sound velocity
in stellar matter with $\Lambda_{1.4}$ and $R_{1.4}$, respectively, were
also observed, but at somewhat different densities. A single determination
$\Lambda_M$ or radius $R_M$ of a  NS  of mass $M$ thus allows to
constrain the thermodynamical properties  at three  different distinct
but close densities.

\begin{figure}
\centering
\resizebox{0.75\columnwidth}{!}{%
\includegraphics{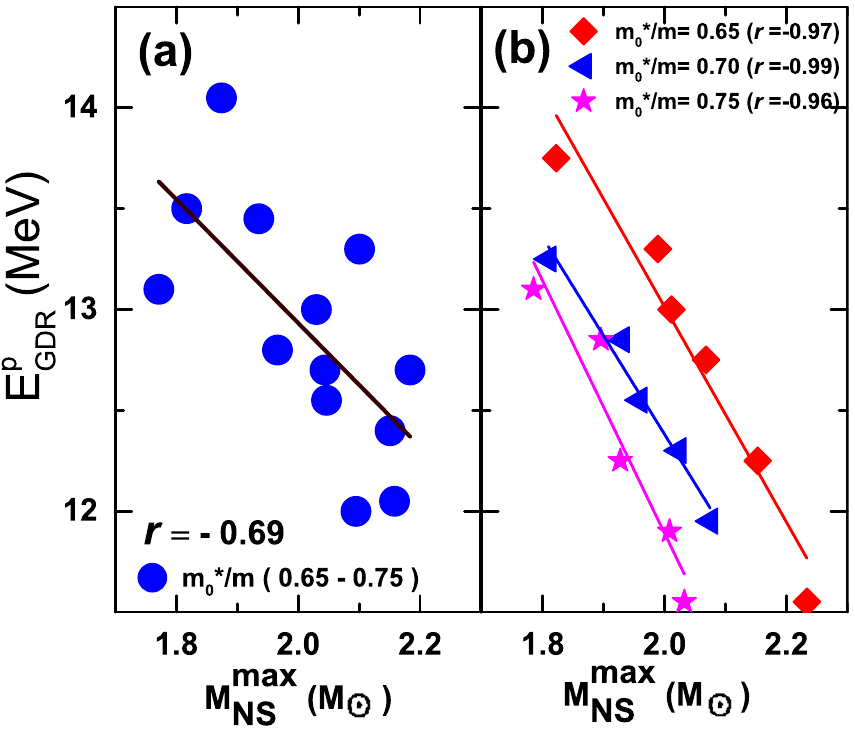} }
\caption{Correlation of $E_{\rm GDR}^{\rm p}$
and $M_{\rm NS}^{\rm max}$ obtained using (a) the set of selected models
as in Fig. \ref{fig:17} with effective mass  $m^*_0/m$ in the range 0.65
-0.75 and (b) a set of systematically varied models with chosen fixed effective masses
in the present work. The figure is taken from Ref. \cite{Malik2019}.}
\label{fig:18}       
\end{figure}

\subsection{In tandem with micro-physics: proposing a new EoS}

The laudatory approach taken in the meta-modeling of the EoS serves 
as a guide to reach to probable values of the thermodynamical variables at
different densities. Its success, however, depends on the precision choice
of the input nuclear matter  parameters.  They may not be beyond question. The
truncation order (see Eqs. (\ref{xeq3}) and (\ref{xeq4})) or the size of
the parameter space may also deem to be a suspect, particularly, at very
low densities $\leq 0.5\rho_0$ and at very high densities, relevant to
massive neutron stars \cite{Ferreira2020}. However, the efficacy of the
EoS in matching many details of microscopic nuclear physics in tandem
with astrophysical observations has not yet been soundly checked, even
if the few input nuclear matter  parameters entering the meta-modeling
might be consistent.

Close inspection of  nuclear matter  EoS shows that it must be stiff
enough to support a NS mass of $\sim 2M_\odot$, but soft enough so
that $\Lambda_{1.4} < 580$ \cite{Abbott2017}. Attempts have been made to
establish a connection between the tidal deformability  and microscopic
nuclear physics from different ends, from sophisticated  microscopic
modeling of the low density EoS in chiral effective field theory (CEFT)
\cite{Tews2018,Lim2018,Tews2019,Lim2019} or from use of a RMF inspired family
of EoS models calibrated to provide a good description of a set of finite
nuclear properties \cite{Fattoyev2018}. The somewhat ambivalent outcomes
give the realization  that the connection of $\Lambda$ to the laboratory
data is not yet fully transparent and that more stringent constraints on
the isovector sector of the effective interaction are needed.

The  robust correlations of $\Lambda_{1.4}$ and $R_{1.4}$ with selective
linear combinations of isoscalar and isovector properties of nuclear
matter \cite{Malik2018} throw a strong hint that 
the isovector giant resonances in conjunction with the
isoscalar resonances in finite nuclei may help in guiding to such
strong constraints. To have a deeper look into these
nuances of microscopic nuclear physics, Malik {\it et al} \cite{Malik2019}
chose to study few experimental data of particular interest involving
isoscalar and isovector properties of finite nuclei, namely, (i) the
centroid energy $E_{\scriptsize{GMR}}$ of isoscalar giant monopole resonance (ISGMR),
(ii) the peak energy $E_{GDR}^p$ of the isovector giant dipole resonance
(IVGDR), and (iii) the dipole polarizability $\alpha_D$, all for the
heavy nucleus $^{208}$Pb, corroborated with results from astrophysical
sector, namely, the maximum mass of a NS and the tidal deformability. The
analysis was done in the Skyrme framework with twenty-eight chosen 'best'
accepted  Skyrme EDFs.    These EDFs provided a satisfactory
reproduction of the binding energies of finite nuclei and their charge
radii, and obeyed reasonable constraints on the saturation density
($\rho_0 = 0.16 \pm 0.01$ fm$^{-3}$), binding energy ($e_0 = -15.8\pm
0.5$ MeV), isoscalar nucleon effective mass ($m_0^*/m = 0.60 - 1.00$)
and the isoscalar nuclear matter  incompressibility ($K_0  = 240\pm 30$ MeV).
The results, displayed in Fig. \ref{fig:17}, with the maximum mass of a
NS $M_{\rm max}$ plotted against $\Lambda_{1.4}$ are self explanatory
with the symbols marked there. They show that only 3 of the  28 EDFs
satisfy all the constraints imposed from finite nuclei and astrophysical
data. Commensurate with the constraint on the maximum mass of a NS
($M_{\rm max}\sim 2.0M_\odot$), the tidal deformability for a canonical
star comes out to be $\Lambda_{1.4}\sim 290 - 330$, the effective mass
$m_0^*/m \sim 0.7$ and the isovector mass splitting $\Delta m_0^*\sim
-0.2\delta$. The essence of the survey is that fulfilling simultaneously
isoscalar and isovector constraints with the astrophysical ones is
extremely restrictive.

The conclusion drawn from Fig. \ref{fig:17} on  the tidal deformability
$\Lambda_{1.4}$ is indicative of the domain in which its value may
lie. To have a more quantitative assessment on it, a new EoS , albeit in
the Skyrme framework, has been proposed \cite{Malik2019} with a wider
fit data base. These constraints include the observed maximum NS mass,
the binding energies of  spherical magic nuclei, their charge radii,
the ISGMR energies of $^{90}$Zr, $^{120}$Sn and $^{208}$Pb and the
dipole polarizability $\alpha_D$  of $^{48}$Ca, $^{68}$Ni, $^{120}$Sn
and $^{208}$Pb.  The IVGDR peak energies are left out of the fitting
protocol deliberately. Simultaneously constraining all data impose severe
restrictions on the model parameters. As example, calculations with
the selected EDFs of Fig. \ref{fig:17} reveal the existence of some
anti-correlation ($r\sim -0.69$) of $E_{GDR}^p$ ($^{208}$Pb) with $M_{\rm
max}$ when the EDFs are sorted out in groups  within narrow windows of
$m_0^*/m$ (see panel (a) of Fig. \ref{fig:18}).  This correlation shoots
up to nearly unity ($r\sim 1.0$) when calculated with systematically
varied models with fixed values of $m_0^*/m$ as displayed in panel (b)
of the same figure. For given values of $M_{\rm max}$ and $m_0^*/m$,
$E_{GDR}^p$ is the outcome of the calculation keeping all other data in
the fitting protocol unchanged.

Looking at the trends depicted by Figs.  \ref{fig:17} and
\ref{fig:18}, it is evident that the values of $\Lambda_{1.4}$
is sensitive to the maximum neutron mass as well as various ground
and excited state properties of the finite nuclei. The optimized
$\chi^2$ -function obtained by fitting these data yields the EDF
parameters as  listed in Table \ref{tab5} \cite{Malik2019}. The errors
on the parameters are calculated within the covariance method
\cite{Zhang2018a,Dobaczewski2014}. The central value of
$\Lambda_{1.4}$ comes out to be 267; we call this EDF SK$\Lambda$267. With
this EDF it is seen that all the nuclear matter parameters and the
lower limit on $M_{\rm max}$ are in comfortably acceptable bounds;
$\Lambda_{1.4}$ and $R_{1.4}$ are seen to be in very good agreement
with those reported recently \cite{Sabatucci2020,Bonnard2020}. The value
of $M_{\rm max}$ is (2.04$\pm$0.15)$M_\odot$, that of the neutron skin
$\Delta r_{np}$ for $^{208}$Pb is $0.15\pm 0.05$ fm.

The experimental value of tidal deformability is not settled yet, we
therefore test the tolerance of the fit of the calculated observables with
data by arbitrarily constraining $\Lambda_{1.4}$ to different values. As
example, an extra constraint $\Lambda_{1.4} = 500\pm 100$ was used in the
fit. The outcome is $\Lambda_{1.4}=484\pm 215$, with somewhat different
Skyrme parameters. This EoS, is called SK$\Lambda$484 \cite{Malik2019}
is somewhat stiffer than SK$\Lambda$267. As expected, it produces a larger
$R_{1.4}$ 13.1$\pm$1.4 km compared to 11.6$\pm$1.0 km for a canonical NS,
a larger $\Delta r_{np}$ 0.21$\pm$0.04 fm for $^{208}$Pb and a somewhat
larger $M_{\rm max}$ 2.10$\pm$ 0.04 $M_\odot$. The isoscalar nuclear
matter  parameters like $e_0, K_0, Q_0$ or $m_0^*$ are nearly unaffected,
but the isovector parameters suffer some changes. Overall, it appears ,
the softer model SK$\Lambda$267 is more compatible with the measured
properties of finite nuclei.

\begin{table}
  \caption{List of fit data corresponding to the symmetric nuclear matter
(SNM), pure neutron matter (PNM) and symmetry energy coefficient (SYM)
together with the range of densities in which they are determined.
Here $P(\rho)$ represents pressure of nuclear matter, ${e}_n(\rho)$ is the energy
per neutron in PNM and $C_2(\rho)$ is the symmetry energy coefficient.}
  \label{tab1}
      \setlength{\tabcolsep}{16pt}
      \renewcommand{\arraystretch}{1.1}
     \begin{tabular}{ l c c c c }
      \hline \hline 
           & Quantity & Density region & Band/Range & Ref.  \\ 
            &     & fm$^{-3}$ & (MeV) & \\
      \hline  
      SNM & $P(\rho)$ 	   &  $0.32~{\rm to}~0.74$   &  HIC        &
\cite{Danielewicz2002}  \\
      SNM & $P(\rho)$     &  $0.19~{\rm to}~0.33$   &  Kaon exp       &
\cite{Fuchs2006,Fantina2014}  \\
      & & & & \\
      PNM & ${e}_n(\rho)$  & 0.1            &   $10.9\pm0.5$
& \cite{Brown2013} \\
      PNM & ${e}_{n}(\rho)$   & $0.03~{\rm to}~0.17$   &
N$^{3}$LO        & \cite{Hebeler2013} \\
      PNM & $P(\rho)$            & $0.32~{\rm to}~0.73$   &   HIC
& \cite{Danielewicz2002} \\
      PNM & $P(\rho)$            & $0.03~{\rm to}~0.17$   &   N$^{3}$LO
& \cite{Hebeler2013} \\
      & & & & \\
      SYM & $C_2(\rho)$       &   0.1              &   $24.1\pm0.8$    &
\cite{Trippa2008}\\
      SYM & $C_2(\rho)$       & $0.01~{\rm to}~0.19$       &   IAS,HIC
& \cite{Danielewicz2014,Tsang2009} \\
      SYM & $C_2(\rho)$       &   $0.01~{\rm to}~0.31$      &   ASY-EoS
& \cite{Russotto2016}\\
\hline \hline 
     \end{tabular}
\end{table}

\begin{table}
   \caption{The final model parameters obtained by optimizing the
$\chi^2$ function together with the uncorrelated and correlated errors
(see text for details).  The parameters  $K_1$ and $K_2$ are in units
of MeV.fm$^{-3}$, $a$ and  $b$ are in MeV and $k_+$ and $k_-$ are in
fm$^{3}$.}

  \label{tab2}
\setlength{\tabcolsep}{13pt}
\renewcommand{\arraystretch}{1.1}
\begin{tabular}{ccccccc}
\hline \hline 
 ${\tilde \alpha}$ & $K_1$ & $K_2$ & $a$ & $b$ & $k_+$   &$k_-$  \\
	  \hline
  1.11 & -1220.21 & 977.94 & 120.03 &  -121.93 & 6.07  & 2.60  \\
Unc. err. & 1.16     & 2.38       & 0.15          &  0.33         &   0.10
&    0.15       \\
          Cor. err. &  103.04    &  90.25       &  15.01     &  13.57
&  1.13       &  0.96      \\
\hline \hline 
 \end{tabular}
\end{table}

\begin{table}
   \caption{Different properties pertaining to nuclear matter (NM) and
            neutron star (NS) obtained with the final parameters
            listed in Table \ref{tab2}. $K_{\tau} $  is the symmetry
            incompressibility at saturation density corresponding to ANM:
            it is defined as  $ K_{\tau} = K_{\rm sym}^0 - 6L_0 - Q_0 L_0
            /K_0$ }

   \label{tab3}
    \setlength{\tabcolsep}{28pt}
      \renewcommand{\arraystretch}{1.1}
    \begin{tabular}{lccc}
\hline \hline 
 Type	& & Unit &Value \\
\hline 
  NM &	$e_0$   & MeV			& $-15.93\pm0.20$  	\\
     &	$\rho_0$ & fm$^{-3}$		& $0.1620\pm0.003$	\\
     &	$K_0$ 	& MeV			& $225.23\pm6.50$	\\
     &	$m_0^*/m$ & 			& $0.67\pm0.04$	\\
     &  ${m_{v,0}^*}/{m}$ &         & $0.78 \pm^{0.05}_{0.04}$\\
     &	$\Delta m_0^*/ \delta$ &	& $-0.19\pm0.08$	\\
     &  $C_2(\rho_{0})$  & MeV 		& $33.94\pm0.50$	\\
     &	$L_0$   & MeV			& $68.50\pm3.74$	\\
     &	$K_{\rm sym}^0$ & MeV		& $-47.46\pm17.28$	\\
     &	$K_\tau$ & MeV			& $-349.22\pm13.19$	\\ 
     &	$M_{\rm c}$   & MeV		& $998.79\pm41.38$	\\ 
     &  $Q_0$   & MeV			& $-359.23\pm23.82$	\\
 
     \hline	
  NS  &	$M_{\rm max}^{\rm NS}$ & $M_\odot$	& $2.07\pm0.03$	\\
      & $R_{1.4}$ & km	& $12.63\pm0.17$	  \\
\hline \hline 
  \end{tabular}
  \end{table}

\begin{table}
\caption{The empirical values of $M_0$ and $K_{\rm sym}^0$ derived
for different limits on $\Lambda_{1.4}$ and $L_0$. The value of
$\Lambda<800(400)$ is derived with 90\%(50\%) confidence limit for
GW170817.  The ranges of $L_0=30-86$ MeV and $40-62$ MeV are taken from
references \cite{Oertel2017,Lattimer2013}. The value of $K_{\rm sym}^0$
has a good overlap with that obtained recently \cite{Zimmerman2020}
from NICER and LIGO/VIRGO constraints.  } \label{tab4}

\begin{center}
\setlength{\tabcolsep}{28pt}
      \renewcommand{\arraystretch}{1.1}
\begin{tabular}{lccc}
\hline \hline 
 $L_0$ & $\Lambda_{1.4}$ & $M_0$ & $K_{\rm sym}^0$\\
(MeV) &  & (MeV) & (MeV)\\
\hline
30 – 86  & 0 – 800    & 1972 – 2878     &  -143.8 – 19.8      \\
         & 0 – 400    & 1972 – 2042     &  -143.8 – 17.3      \\
40 – 62  & 0 – 800    & 1836 – 3206     &  -115.5 – -48.2       \\
         & 0 – 400    & 1836 – 2371     &  -115.5 – -50.7       \\
\hline \hline 
\end{tabular}
\end{center}
\end{table}

\begin{table*}[t]
\caption{Parameters for the  model  Sk$\Lambda$267 and the resulting
nuclear matter and neutron star properties along with  their errors
in the parenthesis. $C_2^0$ is the symmetry energy coefficient, $L_0$
is related to its density derivative \cite{Malik2018a}. } \label{tab5}
\setlength{\tabcolsep}{0.4pt}
\centering\begin{tabular}{cccccccccc}
\hline \hline 
\rule{0pt}{3ex}%
$ t_0$&  $ t_1$ & $t_2$ &$ t_3$
 & $x_0$ &$ x_1$ &$ x_2$ &$ x_3 $& $\alpha$ &
$W_0$ \\
 ( MeVfm$^3$ ) & ( MeVfm$^5$ ) & ( MeVfm$^5$ ) &
( MeVfm$^{3+3 \alpha}$ ) &  & & &&  &
( MeVfm$^5$ ) \\
\hline
 -2481.08 & 482.51 & -516.17 & 13778.74 & 0.93 & -0.53 & -0.97 & 1.54 & 0.167 & 121.38  \\
 (89.05) & (50.41) & (407.22) & (123.72) & (0.28) & (0.89) & (0.20) &  (0.58) & (0.018) &  (9.35) \\
\hline
 $e_0$& $\rho_0$& $K_0$ & $m_0^*/m $ &$C_2^0$
&$L_0$&$\Delta m_0^*/\delta$& $\Lambda_{1.4}$ & $R_{1.4}$  & $M_{\rm NS}^{\rm
 max}$ \\
 (MeV) & (fm$^{-3}) $& (MeV) &  &(MeV)
&(MeV)& &  &  (km) &  ($M_\odot$) \\
\hline
 $16.04$& 0.162& 230.2& 0.70&31.4& 41.1& -0.25&267 & 11.6& 2.04\\
( 0.20 )&  (0.002) &(6.4) & (0.05)&( 3.1) &( 18.2)& (0.35)& (144)&( 1.0) &(0.15)\\
\hline \hline 
\end{tabular}
\end{table*}

\clearpage

\begin{table}[]
\centering
 \rotatebox{90}{%
   \begin{varwidth}{\textheight}
\caption{\label{tab6} The constraints on isoscalar and isovector
nuclear matter parameters together with  radius and tidal deformability
of neutron star with canonical mass and the neutron-skin thickness in
$^{208}$Pb nucleus derived from microscopic and macroscopic data in the
present work. The isoscalar parameters considered are the nuclear matter
incompressibility coefficient $K_0$, skewness parameter $Q_0$ and nucleon
effective mass $m_0^*$ and those corresponding to the isovector sector
are the symmetry energy coefficient $C_2^0$, slope $L_0$, curvature
$K_{\rm sym}^0$ and skewness $Q_{\rm sym}^0$ and the  effective nucleon
mass splitting $\Delta m_0^*/\delta$.}

\setlength{\tabcolsep}{6pt}
      \renewcommand{\arraystretch}{2}
\begin{tabular}{ccccccccc}
\hline \hline 
\multicolumn{2}{c}{\multirow{2}{*}{Quantity}}      & \multirow{2}{*}{Unit} & \multicolumn{6}{c}{Values from Refs.}                                                                                                                                                            \\ \cline{4-9} 
\multicolumn{2}{c}{}                               &                        & \cite{Mondal2017} & \cite{Malik2018a} & \cite{Malik2019} & \cite{RocaMaza2013a}  & \cite{De2015} & \cite{Agrawal2012,Agrawal2013} \\ \hline
\multirow{3}{*}{SNM} & $K_0$                       & \multirow{2}{*}{MeV}   &                          & $225.0\pm6.4$            & $230.0\pm6.4$            &                                               & $212.0\pm20.1$           &                          \\
                     & $Q_0$                      &   &                          & $-359.0\pm23.0$           &
$-367.0\pm12.0$           &                                               & $-378.0\pm31.6$           &                          \\
                     & $m_0^{\star}/m$             &  -                      & $0.7\pm0.1$              & $0.68\pm0.04$            & $0.70\pm0.05$            &                                               &                          &                          \\ \hline
\multirow{5}{*}{ANM} & $C_2^0$                       & \multirow{4}{*}{MeV}   & $32.1\pm0.31$            & $33.94\pm0.5$            & $31.4\pm3.1$             & $31 \pm 2_{\rm est}$                              &                          &                          \\
                     & $L_0$                       &                        & $60.3\pm14.5$            & $68.5\pm3.74$            & $41.1\pm18.2$            & $43 \pm 6_{\rm exp} \pm 8_{\rm the} \pm 12_{\rm est}$ &                          & $59.3\pm12.8$            \\
                     & $K_{\rm sym,0}$             &                        & $-111.8\pm71.3$          & $-47.46\pm17.28$         & $-123.9\pm70.2$          &                                               &                          &                          \\
                     & $Q_{\rm sym,0}$             &                        & $296.8\pm73.6$           & $394.07\pm24.96$                      & $564.63\pm98.73$         &                                               &                          &                          \\
                     & $\Delta m_0^{\star}/\delta$ &    -                    & $0.17\pm0.24$            & $-0.19\pm0.08$           & $0.25\pm0.35$            &                                               &                          &                          \\ \hline
$^{208}$ Pb  & $\Delta r_{np}$     & fm    &    &      &      &                                               &                          & $0.195\pm0.022$          \\ \hline
\multirow{2}{*}{NS}  & $R_{1.4}$                   & km                     &                          &     $12.63\pm0.17$                     &                      $11.6\pm1.0$    &                                   &                          &                          \\
                     & $\Lambda_{1.4}$             &  -
&                          &                          & $267 \pm 144$
&     &                          &                                              \\ 
\hline \hline 
\end{tabular}
\end{varwidth}}
\end{table}

In Table \ref{tab6}, we collect the values of various isoscalar and
isovector nuclear matter parameters constrained by microscopic and
macroscopic data considered in the present work.  We also provide the
constraints  on the radius and tidal deformability for the neutron
star with the canonical mass and the neutron-skin thickness in the
$^{208}$Pb nucleus. This table is far from being complete and may be
complemented with the constraints  provided recently, for instance
Ref. \cite{RocaMaza2018}.  The splitting of effective nucleon mass are
found to be positive as well as negative. The negative values are obtained
only when the electric dipole polarizability in nuclei and the maximum
mass of the neutron star are constrained simultaneously.  The  values
of $L_0$ and $\Delta r_{np}$ presented in the last column of the table
are obtained by combining the results from three  different ansatz for
the density dependence of the symmetry energy.
{\bl The best fit  estimates for radius and the mass of the  PSR
J0030+0451 obtained by NICER are $R=13.02^{+1.24}_{-1.06}$ km for the
$M_{\rm NS} = 1.44^{+0.15}_{-0.14} M_\odot$ \cite{Miller2019}. These
values have good overlap  with the ones presented in the Table \ref{tab6}.
More precise values of these quantities are , however,  required to
constrain the nuclear matter parameters and the EOS.}

\section{Summary and Outlook } 

All the rich physics of the  interacting nucleons is telescopically
encoded  in the nuclear matter parameters; a model-independent EoS of
symmetric and asymmetric nuclear matter can be built on them. A convenient
means to link these parameters to the properties of finite nuclei and
of neutron stars is provided by the nuclear mean-field models. The
parameters such as the binding energy per nucleon, the saturation
density, the nuclear matter  incompressibility coefficient and symmetry energy
coefficient have been determined within a narrow window from the ground
state and excited state properties of finite nuclei, the higher order
density derivatives such as the density slope of symmetry energy or the
skewness parameter are still not known in comparative precision. Attempts
have been made in the last few years to contain them in narrow bounds
through correlation analysis, we review them in this article. The Skyrme
mean-field approach, till date has proved to be the most comprehensive in
explaining diverse nuclear data, the review  places more emphasis on it.

The nuclear matter  parameters reflect different properties of nuclear matter,
but they may be correlated to each other being the underpinnings of
the varied aspects of the same nucleonic interactions. We have explored
this correlation property analytically in the mean-field approach. The
better known low order density derivatives help in narrowing down the
uncertainty in the higher order derivatives.  Additional information
comes from the correlations of selective nuclear matter  parameters with selective
nuclear observables. For example, the symmetry density slope is sensitive
to the neutron-skin of a finite nucleus as well as to the radii and
the tidal deformability of neutron stars. Neutron star properties have
been seen to be particularly sensitive to the higher order nuclear matter 
parameters. Present state of art technology, however, has not been able
to contain the data emanating from neutron stars in more tight bounds.
We hope, from the NICER mission and from the future advanced LIGO-VIRGO
detection of gravitational waves, the cosmic data may find more fine
resolution and their consonance with laboratory micro-physics may guide
to a better understanding of the nuclear matter  equation of state.

\clearpage

\pagebreak
{\bf Acknowledgments} 

The authors acknowledge the  the contribution  of many collaborators,
who over  many years were instrumental in helping  to develop the ideas
that we threaded in this review.  The authors are extremely thankful to
Tanuja Agrawal for her assistance in the preparation of the manuscript.
T. M. acknowledges the hospitality extended to him by Saha Institute of
Nuclear Physics during the course of this work.  J. N. D. acknowledges
support from the Department of Science and Technology, Government of
India with grant no. EMR/2016/001512.

\clearpage

\pagebreak

\bibliographystyle{spphys.bst}

\end{document}